\documentclass[12pt,dvips,a4paper]{article}
\usepackage[pdftex]{graphicx,xcolor}

\usepackage{amsmath,amssymb,graphicx,array}

\usepackage{setspace}
\usepackage[round]{natbib}
\RequirePackage{lineno}

\newcommand{\ignore}[1]{}

\renewcommand{\title}{3D network modelling of fracture processes in fibre-reinforced geomaterials}

\begin{document}

\begin{center} \begin{LARGE} \textbf{\title} \end{LARGE} \end{center}

\begin{center}
  Peter Grassl$^{*,1}$ and Adrien Antonelli$^{1,2}$ 

  $^{1}$School of Engineering, University of Glasgow, UK

  $^{2}$Ecole Normale Sup\'{e}rieure Paris-Saclay, France
  
$^*$Corresponding author: Email: peter.grassl@glasgow.ac.uk, Phone: +44 141 330 5208

\end{center}

\section*{Abstract}
The width of fracture process zones in geomaterials is commonly assumed to depend on the type of heterogeneity of the material.
Still, very few techniques exist, which link the type of heterogeneity to the width of the fracture process zone.
Here, fracture processes in geomaterials are numerically investigated with structural network approaches, whereby the heterogeneity in the form of large aggregates and low volume fibres is modelled geometrically as poly-dispersed ellipsoids and mono-dispersed line segments, respectively.
The influence of aggregates, fibres and combinations of both on fracture processes in direct tensile tests of periodic cells is investigated.
For all studied heterogeneities, the fracture process zone localises at the start of the softening regime into a rough fracture.
For aggregates, the width of the fracture process zone is greater than for analyses without aggregates.
Fibres also increase the initial width of the fracture process zone and, in addition, result in a widening of this zone due to fibre pull out.

Keywords: fibres, fracture, geomaterial, heterogeneity, roughness

\section{Introduction}
Many structures made of geomaterials exhibit failure processes which are influenced by the heterogeneity of the material at an intermediate (meso-) scale.
For instance, the type of coarse aggregates in concrete influences stiffness, strength and fracture energy of the material.
For fibre reinforced cementitious materials, fibre type and geometry strongly influence the tail of the stress-crack opening curve \citep{NaaNamAlw91,LiWu92}.
Therefore, modelling approaches which link the geometry, spatial distribution and mechanical properties of individual constituents at the meso-scale to the structural response are attractive.
Furthermore, detailed investigations based on experiments and computational modelling of the mechanical interaction of individual constituents can contribute to further understanding of failure processes at larger scales.

Numerical approaches based on nonlinear fracture mechanics (NLFM) \citep{Dug60,Bar62} are commonly used to predict the failure of structural components of practical size, since the length of the fracture process zone is too large (with respect to the size of the structural component) for linear elastic fracture mechanics (LEFM), but too small for plastic limit load analysis to be applicable.
Here, fracture process zone is defined as the zone in which energy is dissipated at a certain stage during the fracture process.
Within computational frameworks, such as the finite element method and discrete stiffness approaches, NLFM is applied in the form of cohesive-crack and crack-band models.
In cohesive-crack models, the displacement field across the fracture process zone is replaced by a displacement jump representing the crack opening and stresses are determined from a stress-crack opening law \citep{HilModPet76,CarPraLop97}.
In crack-band models, the displacement jumps are transformed into cracking strains, so that the stress is calculated using stress-strain laws taking into account the size of the regions in which strains localise \citep{BazOh83}.
This size is usually a function of the element size, so that the load-displacement curves obtained with this approach are mesh-independent \citep{JirBau12}.
Discrete approaches describe both elastic and inelastic responses by means of force-displacement laws between discrete bodies \citep{SchMie92b,SchMie92c,BolHonYos00}.
Often, these force-displacement laws are chosen to be very similar to crack band approaches \citep{GraBol16}.
These different computational NLFM approaches can model the length of the fracture process zone along the fracture, but not its width.

Continuum mechanics is an alternative to nonlinear fracture mechanics, where the fracture process zone is represented by localised but regular fields of displacements.
This is achieved by including a length parameter in continuum models \citep{PijBaz87,BazJir02}.
Maintaining a regularised displacement field during fracture simulations provides mesh-independent solutions upon mesh refinement.   
However, the length parameter influences the numerically predicted peak load and deformation capacity of structures \citep{XenGra16}.
Therefore, this parameter should be chosen so that the localised field of displacements matches the width of the fracture process zone of the material \citep{XenGreMorGra15}.

The fracture process zone in heterogeneous materials such as concrete has been investigated experimentally and numerically.
Experimental studies for fracture in plain concrete in \cite{MihNomNii91,MihNom96,OtsDat00,GreSolLef14} showed that the fracture process zone consists of a narrow band of high dissipation surrounded by a wider region of low dissipation. Fracture surface measurements were also performed to provide further inside into the link between roughness and fracture behaviour \citep{LanJenSha93,MorMorBou06,MorBonPon08,PonBonAur06}.
In \citet{GraJir10,GraGreSol12,GreSolLef14,XenGreMorGra15}, information about the width of the fracture process zone was determined numerically using two-dimensional structural network approaches for the meso-scale of plain concrete consisting of coarse aggregates embedded in a mortar matrix. Numerical models for fibre reinforced concrete, in which fibres were modelled discretely, were proposed in \citet{BolSai97,LeiSloMih04,LeiSlo07,Kab07,RadSimSlu10,KunOguUed11,SchCus11,CagEtsMar12,MonSchMed13,KanKimLim14,ZhaMes16,KanBol17,MonCifMed17}. Most of these studies on fibre reinforced composites aimed at predicting the influence of fibres on stiffness, strength and ductility. There is less information available on how fibres affect the spatial distribution of dissipated energy at the meso-scale.

The aim of this work is to obtain more information about fracture processes in geomaterials at the meso-scale by using a three-dimensional structural network model.
The meso-structure of geomaterials is idealised to consist of a matrix with coarse aggregates, interfacial transition zones (ITZs) between matrix and aggregates, and fibres.
Periodic direct tension analysis are performed and the effect of aggregates and fibres on the stress-displacement curves and spatial distribution of energy dissipation are investigated. 

\section{Method} \label{sec:method}
The present numerical approach for obtaining information on fracture processes in fibre-reinforced quasi-brittle materials relies on periodic meso-structure generation, periodic network modelling of the material response, and roughness evaluation of the fracture patterns obtained from the network modelling. In the following sections, the individual modelling techniques are described in more detail.

\subsection{Periodic meso-structure generation} \label{sec:meso}
\begin{figure}
  \begin{center}
    \begin{tabular}{cc}
      \includegraphics[width=7.5cm]{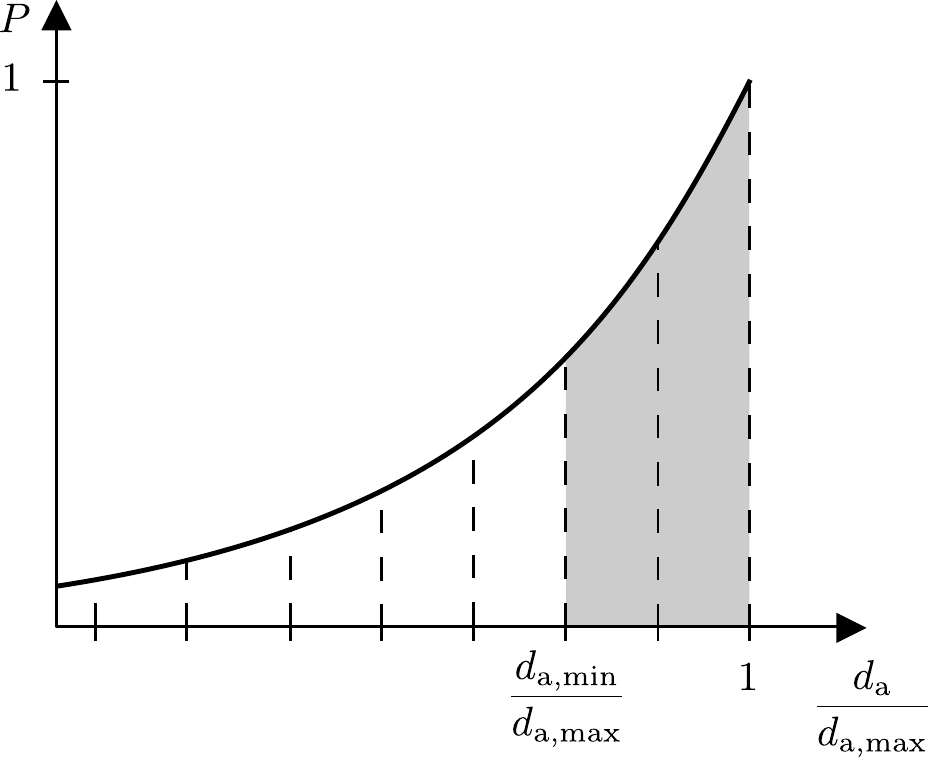} & \includegraphics[width=5cm]{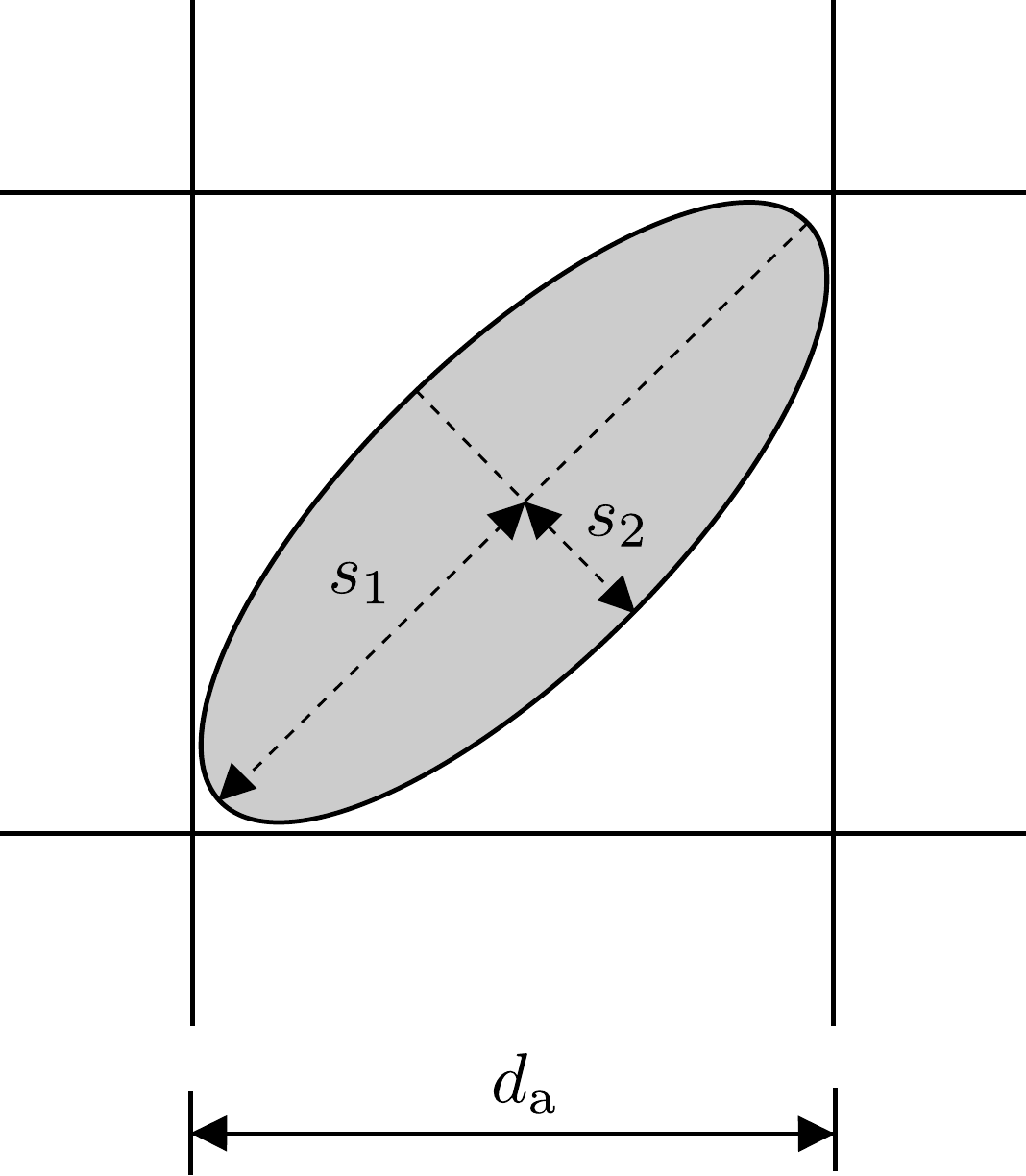}\\
      (a) & (b)\\
    \end{tabular}
    \caption{Generation of poly-dispersed ellipsoids (a) Sieve curve based on Fuller's curve and (b) geometrical restriction imposed by square sieve size.}
    \label{fig:shapeGeneration}
  \end{center}
\end{figure}

The meso-structure of concrete is modelled as coarse aggregates and fibres embedded in a mortar matrix.
Aggregates and fibres are idealised as poly-dispersed ellipsoids and mono-dispersed line segments, respectively.
They are periodically arranged in a computational cell representing the meso-structure of the material.
For a given volume fraction of ellipsoids, Fuller's grading curve is used to determine the size distribution of ellipsoids (Figure~\ref{fig:shapeGeneration}a).
The total volume of ellipsoids is divided into intervals using sieve sizes. For each volume interval, the upper and lower sieve sizes are $m$ and $n=m/2$, respectively.
Here, $m$ is smaller than or equal to the maximum sieve size $d_{\rm a,max}$ and $n$ is greater than or equal to the minimum sieve size $d_{\rm a,min}$.
Starting with the volume interval obtained with the largest pair of sieve sizes, ellipsoids are generated randomly with radii $s_3>s_2>s_1$ so that they fit through the square sieve size $m$, but not $n$ (Figure~\ref{fig:shapeGeneration}b) as proposed in \cite{SloLei99,LeiSlo07} and further investigated in \cite{Meh11}.
This results in the conditions
  \begin{equation}\label{eq:s1}
    \dfrac{1}{2}\sqrt{\dfrac{2}{r^2+1}}r n <s_{1}<\dfrac{1}{2}m  
  \end{equation}
  \begin{equation}\label{eq:s3}
    s_3 = s_1/r
  \end{equation}
  \begin{equation}\label{eq:s2}
    \max\left(s_1, \sqrt{\frac{n^2}{2}-s_1^2}\right) \leq s_2 \leq \min \left(s_3,\sqrt{\frac{m^2}{2}-s_1^2}\right)
\end{equation}
Here, $r$ is uniformly distributed between 0.5 and 1.
Furthermore, $s_1$ and $s_2$ are uniformly distributed between the limits stated in (\ref{eq:s1})~and~(\ref{eq:s2}), respectively.
Line segments are assumed to be of uniform length $l_{\rm f}$. For a volume fraction $\rho_f$, the number of fibres are calculated as $n_{\rm f} = 4 \rho_{\rm f} V/(\pi d_{\rm f}^2 l_{\rm f})$, where $V$ is the volume of the unit cell and $d_{\rm f}$ is the diameter of the fibres.

The input parameters for the meso-structure generation are the volume fraction of ellipsoids $\rho_a$, the maximum and minimum sieve sizes $d_{\rm a,max}$ and $d_{\rm a,min}$, respectively, the volume fraction of line segments $\rho_{\rm f}$, fibre length $l_{\rm f}$ and the diameter of fibres $d_{\rm f}$.
Only ellipsoids greater than the sieve size $d_{\rm a,min}$ are generated, as indicated by the shaded region in Figure~\ref{fig:shapeGeneration}a.

\begin{figure}
  \begin{center}
    \begin{tabular}{ccc}
      \includegraphics[height=3cm]{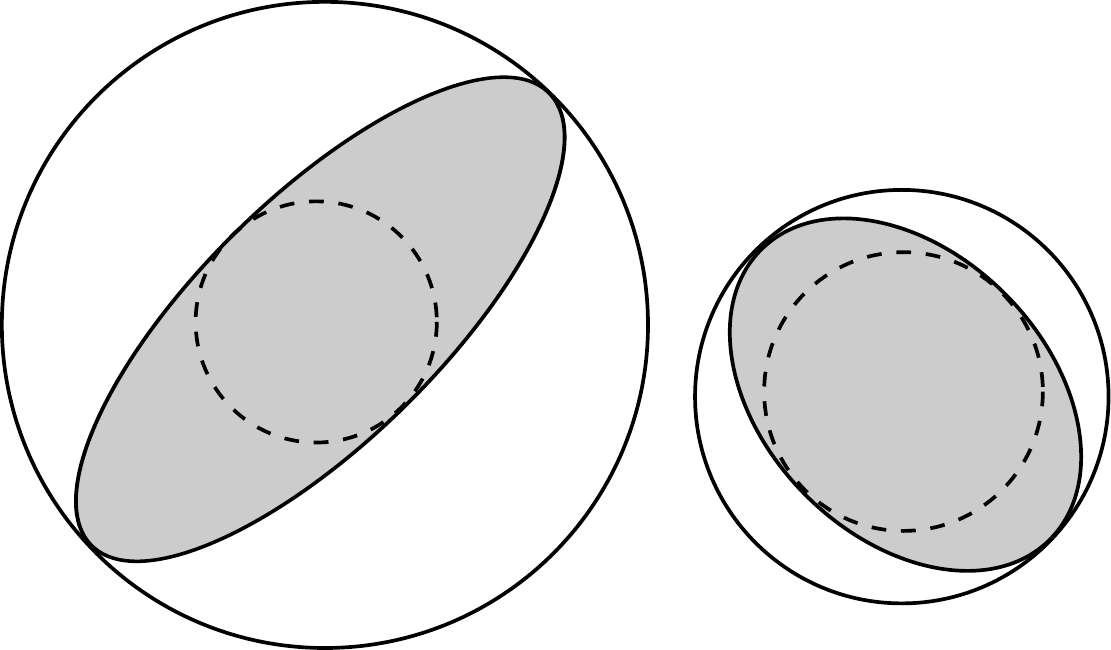} & \includegraphics[height=3cm]{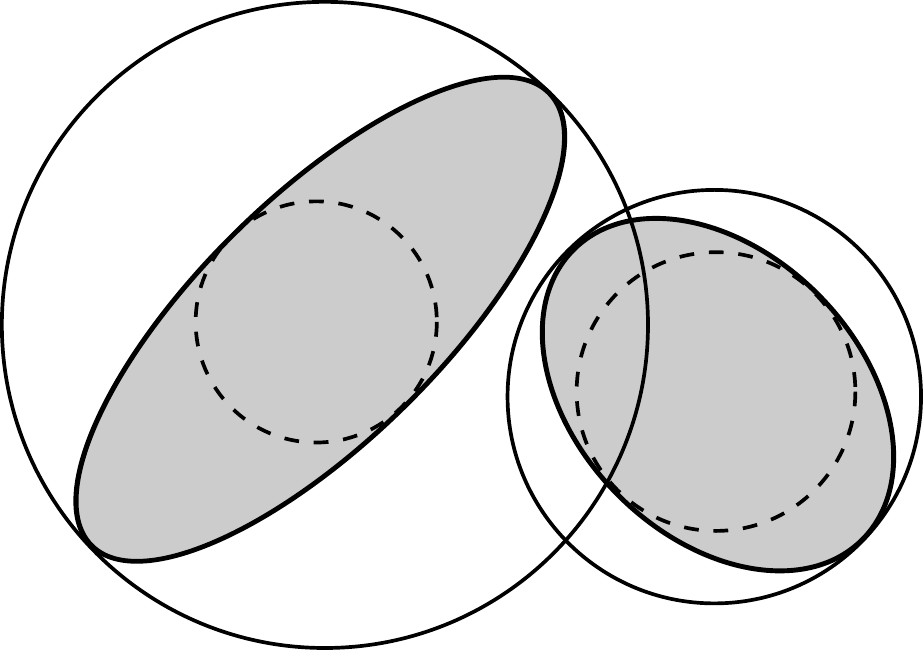} &  \includegraphics[height=3cm]{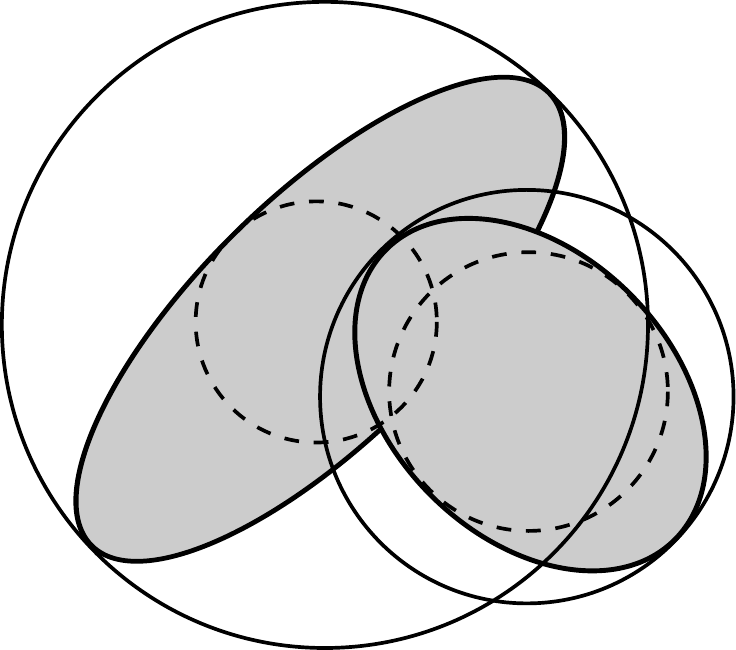}\\
      (a) & (b) & (c)
\end{tabular}
    \caption{Overlap check of ellipsoids using bounding spheres.}
    \label{fig:overlapCheck}
  \end{center}
\end{figure}

Next, ellipsoids and line segments are placed in the periodic cell by a random sequential addition approach \citep{Fed80} so that the centroids of ellipsoids and line segments are within the cell.
Attention is paid so that the random orientation of ellipsoids and line segments are uniformly generated within the volume \citep{Mul59}.
For every randomly placed object, overlap with previously placed objects is checked.
If overlap is avoided, the object is placed in the cell and 26 mirror objects in the adjacent cells are generated by shifting the object to the adjacent periodic cells.
If overlap is detected, a new random position and orientation is generated.
This process is repeated until all objects are placed in the cell.
For the overlap check between ellipsoids, the algebraic system of equations in \cite{WanWanKim01} is used (Figure~\ref{fig:overlapCheck}).
Compared to the overlap check for spheres, solving this system of equations is slow.
Therefore, outer and inner bounding spheres of the ellipsoids are used to exclude any unnecessary checks of ellipsoids.
If the outer bounding spheres of two ellipsoids do not overlap, the two ellipsoids themselves do not overlap (Figure~\ref{fig:overlapCheck}a).
If the inner bounding spheres of two ellipsoids overlap, the two ellipsoids overlap (Figure~\ref{fig:overlapCheck}c).
Only if the outer bounding spheres overlap and the inner spheres do not overlap, the overlap check of two ellipsoids is performed (Figure~\ref{fig:overlapCheck}b).
This simple method based on bounding outer and inner spheres requires significantly less time than applying the method in \cite{WanWanKim01} to all ellipsoids.
For combinations of ellipsoids and line segments, only overlaps between ellipsoids, and ellipsoids and line segments are checked.

\begin{figure}[htbp!]
  \begin{center}
    \begin{tabular}{ccc}
    \includegraphics[width=4cm]{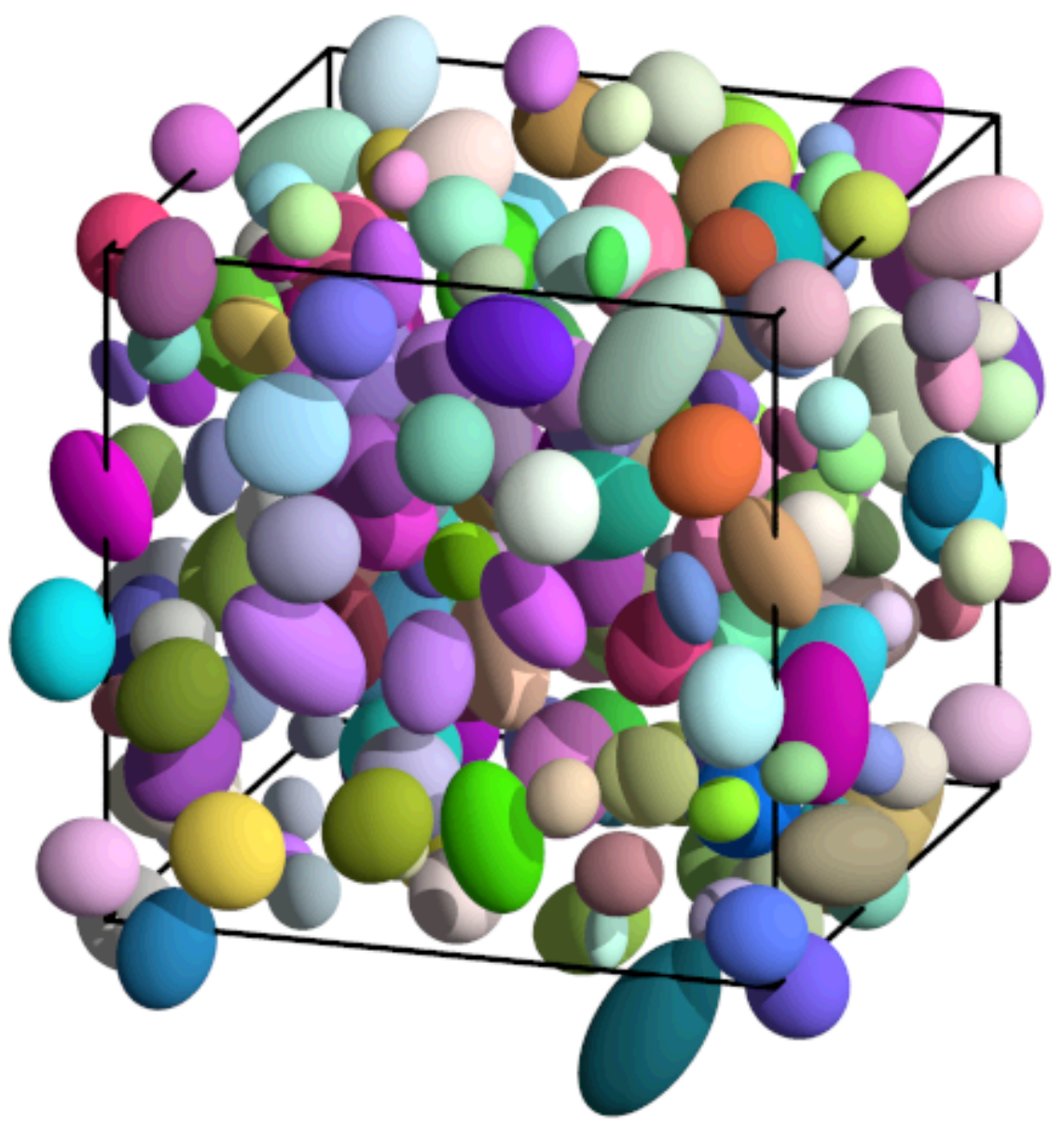} & \includegraphics[width=4.5cm]{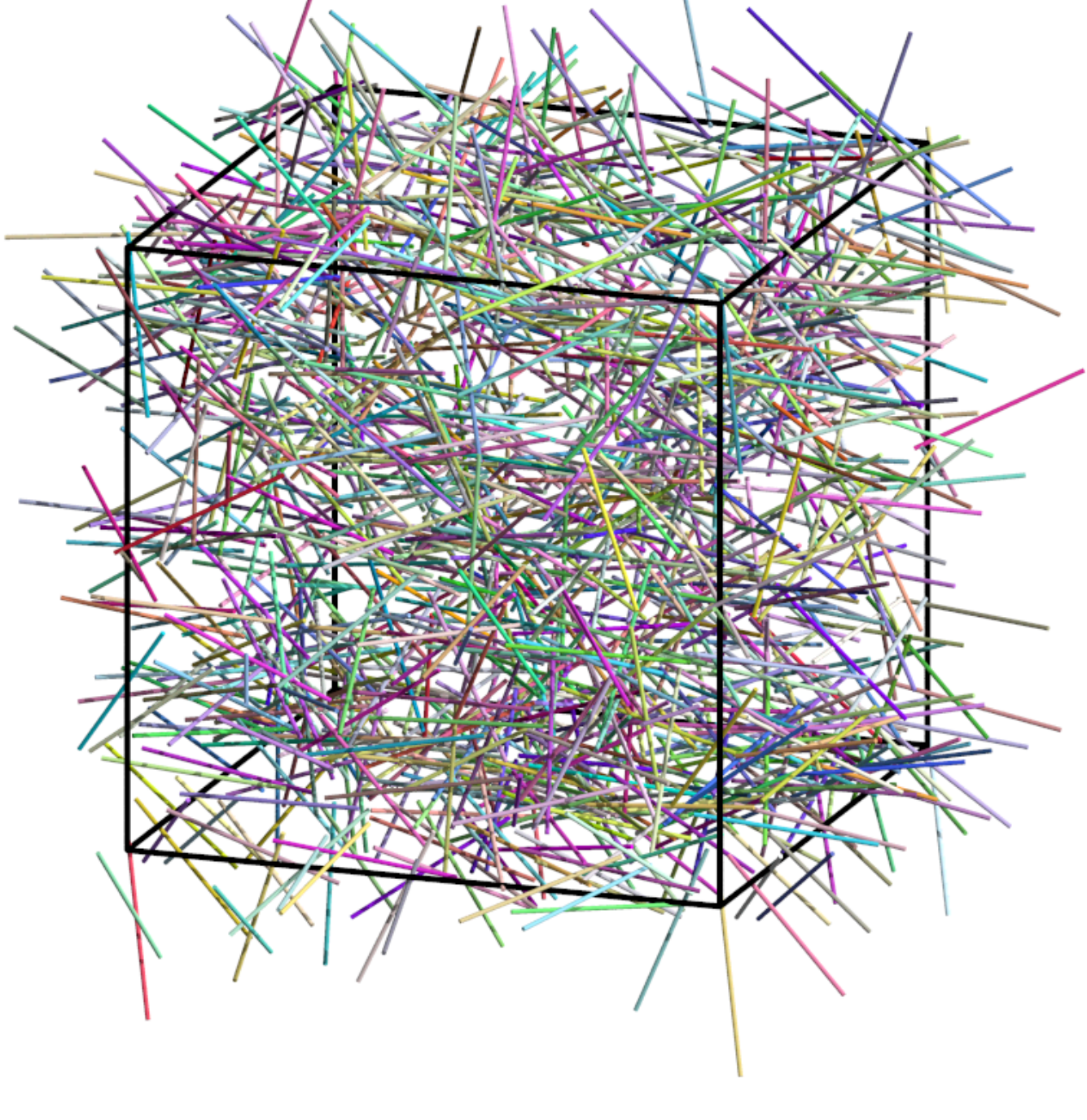} & \includegraphics[width=4.5cm]{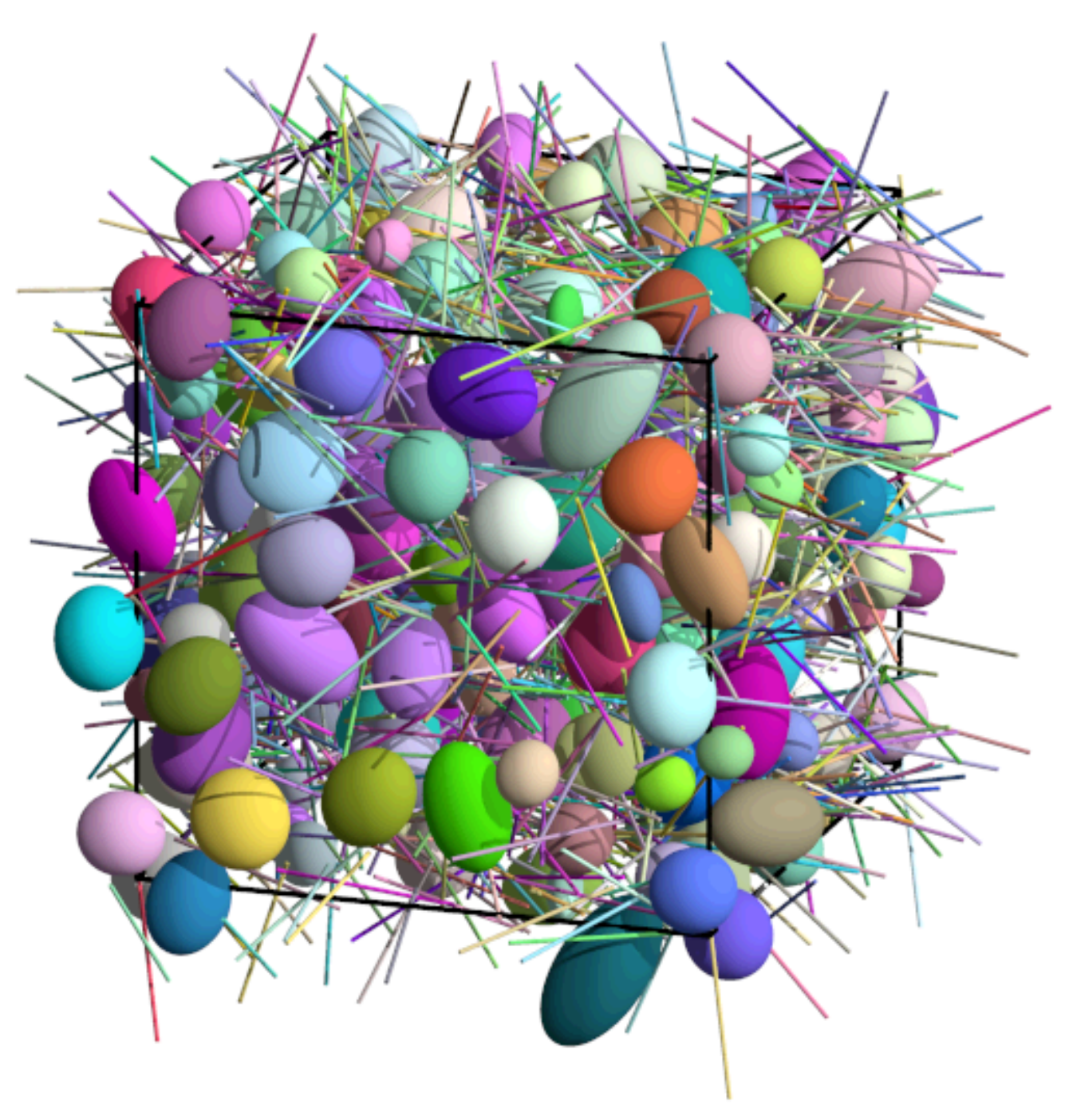}\\
      (a) & (b) & (c)\\
      \end{tabular}
      \caption{Periodic meso-scale generation for (a) ellipsoids, (b) line segments and (c) combination of ellipsoids and line segments.}
      \label{fig:mesoscale}
  \end{center}
\end{figure}

Examples of generations of ellipsoids with $d_{\rm a,max} = 16$~mm, $d_{\rm a,min} = 8$~mm and $\rho_{\rm a} = 0.8$, line segments with $l_{\rm f} = 30$~mm,  $d_{\rm f} = 0.75$~mm and $\rho_{\rm f} = 0.01$, and a combination of ellipsoids and line segments with $d_{\rm a,max} = 16$~mm,  $d_{\rm a,min} = 8$~mm,  $\rho_{\rm a} = 0.8$, $l_{\rm f} = 30$~mm, $d_{\rm f} = 0.75$~mm and $\rho_{\rm f} = 0.01$ are shown in Figure~\ref{fig:mesoscale} for a cell with an edge length of $100$~mm. The fibre diameter $d_{\rm f}$ is only required to calculate the number of line segments to be placed, but not for the placement itself. 
Here, $\rho_{\rm a} = 0.8$ is the total volume fraction of ellipsoids, which is significantly greater than the generated volume fraction of $0.23$ between the sieve sizes $16$~and~$8$~mm.

\subsection{Periodic network modelling} \label{sec:network}
\begin{figure}
  \begin{center}
  \begin{tabular}{cc}
    \includegraphics[width=4.5cm]{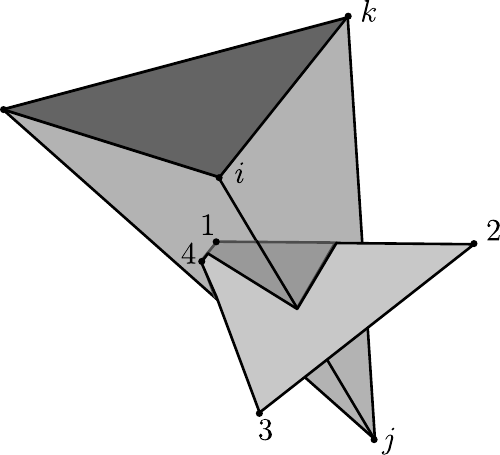} & \includegraphics[width=4.5cm]{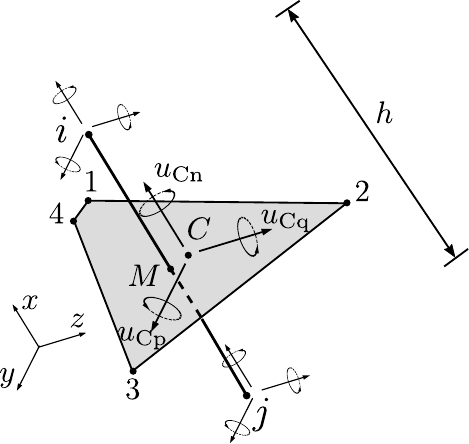} \\
    (a) & (b)
    \end{tabular}
  \caption{3D random network: (a) Example of dual Delaunay Voronoi tessellation and (b) structural element with mid-crosssection.}
  \label{fig:voronoi}
  \end{center}
\end{figure}

The fracture processes at the meso-scale are modelled for a periodic cell subjected to direct tension with a three-dimensional irregular network of discrete structural elements.
The random network generation follows the work in \cite{YipMohBol05}, which was recently extended to dual structural transport problems in \cite{GraBol16}.
For the network generation, random points are placed in the cell using a sequential addition approach enforcing a minimum distance $d_{\rm min}$ between the points \citep{Fed80}. These points are used for dual Delaunay and Voronoi tessellations resulting in randomly arranged tetrahedra and polyhedra.
In Figure~\ref{fig:voronoi}a, one of these tetrahedra with a common facet of polyhedra belonging to two vertices of the tetrahedron is shown.
The network elements are placed on the edges of the tetrahedra. The mid-crosssections of the network elements are set equal to the common facets of the Voronoi cells associated with the element nodes \citep{YipMohBol05}.
The network elements have six degrees of freedom at each node which are linked by rigid body kinematics to displacement jumps at the centroid of the mid-crosssection. These displacement jumps are then related to corresponding stress components using constitutive models described in Section~\ref{sec:constitutive}.  

The information of the spatial arrangement of ellipsoids are mapped onto the network.
According to the position of network elements with respect to ellipsoids, network elements are given the properties of matrix, interfacial transition zone (ITZ) and aggregate. Network elements with both nodes positioned within an ellipsoid are given stiff elastic properties representing aggregates. Elements with both nodes located in the matrix are given properties of mortar with corresponding elastic properties, and strength and fracture energy.
Finally, for elements with one node in an ellipsoid and another one in the matrix or another ellipsoid, the properties of ITZ are used, which are characterised by lower strength and lower fracture energy than those of the matrix. The stiffness of ITZ elements are determined by the harmonic mean of the stiffnesses of matrix and aggregate.

\begin{figure}
  \begin{center}
  \begin{tabular}{cc}
    \includegraphics[width=7cm]{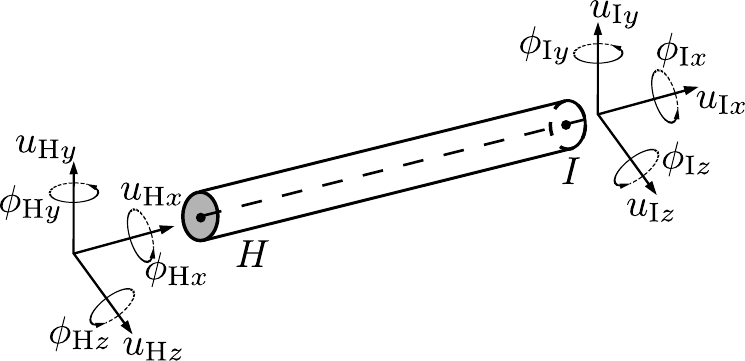} & \includegraphics[width=4.5cm]{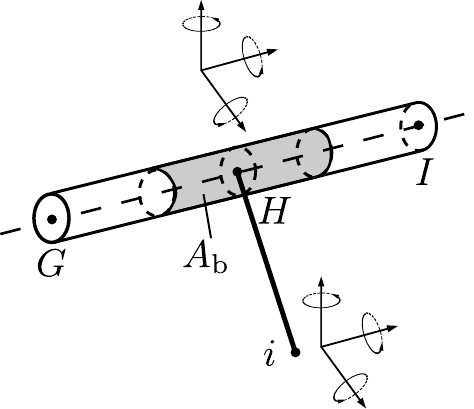} \\
    (a) & (b)
    \end{tabular}
    \caption{Modelling fibres: a) 3D frame element for fibres and b) link element for the interaction between fibres and matrix.}
    \label{fig:fibres}
  \end{center}
\end{figure}

The fibres are idealised as linear elastic structural frame elements \citep{McgGalZie00}, which are placed on the positions of the line segments (Figure~\ref{fig:fibres}a).
Interactions between the frame elements representing fibres and the background network representing matrix and ITZ are modelled by means of link elements as described in \cite{YipMohBol05} (Figure~\ref{fig:fibres}b).
This type of link elements was originally used for the modelling of bond in reinforced concrete \citep{NgoSco67}, and was more recently applied to network models in \cite{BolSai97,MonCifMed17}.
Rigid body kinematics are used to determine, from the nodal degrees of freedom of the link and frame elements, the translation and rotation jumps at the node of the frame element (Figure~\ref{fig:fibres}b). The coordinate system for these jumps is orientated so that one of the axes is aligned with the axial direction of the frame element. For the translation jump in the direction of the frame element, an elasto-plastic model described in Section~\ref{sec:constitutive} is used to model the slip between the frame element and the background network. For the other components, a linear force-displacement law with a 1000 times higher stiffness than the elastic stiffness of the bond law is applied.

Reduction of the embedded length due to pullout of the fibres as discussed in detail in \cite{NaaNamAlw91} is not modelled here, since only small displacements with respect to the pull-out length are considered. Computationally more efficient semi-discrete approaches described in \cite{KanKimLim14,KanBol17} would be well suited to describe the full pull-out process, since these approaches incorporate important features of the fibre-matrix interaction without modelling individual degrees of freedom.

\begin{figure}[htbp!]
  \begin{center}
    \includegraphics[width=10cm]{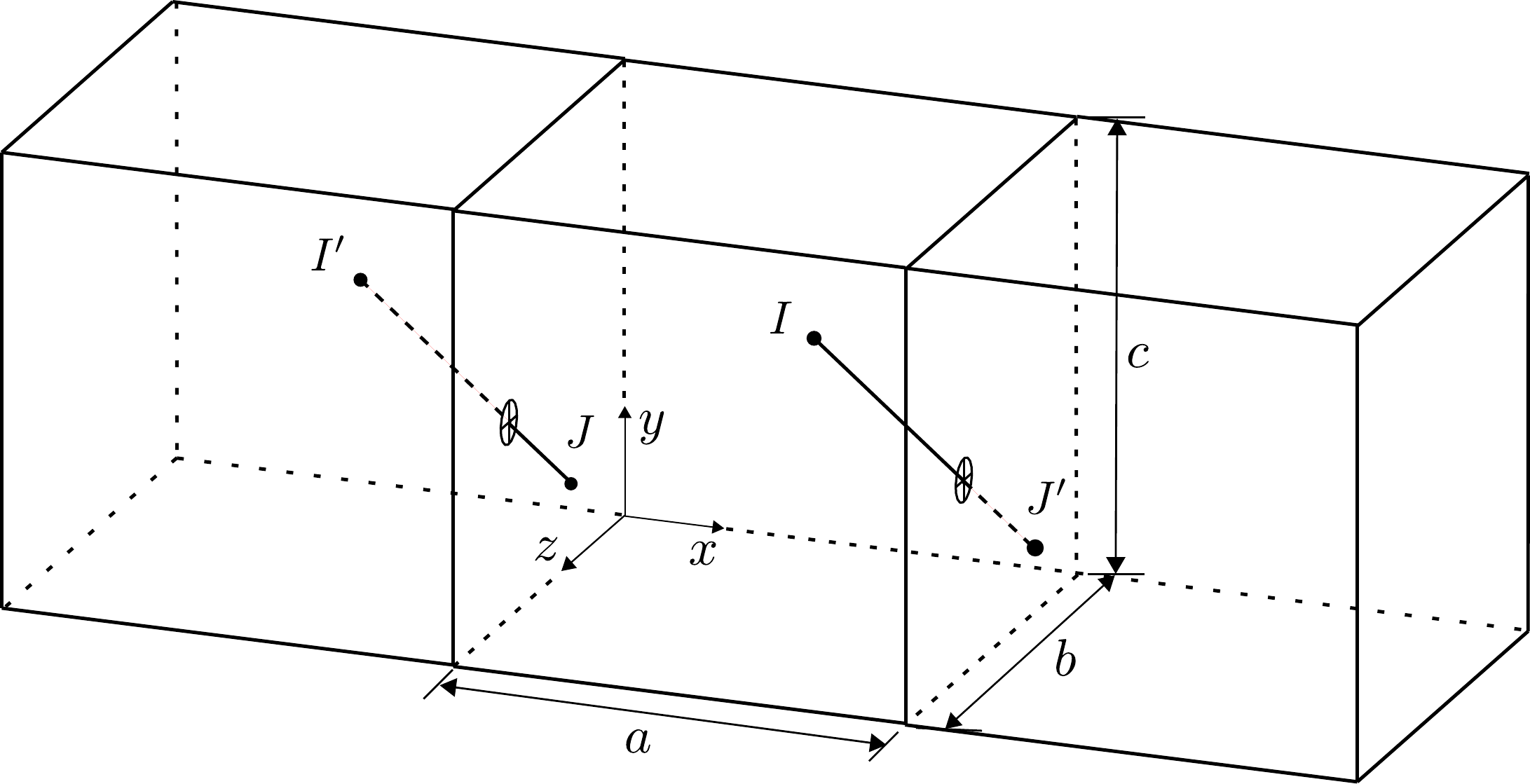}\\
    \caption{Periodic generation of background network.}
    \label{fig:periodic}
  \end{center}
\end{figure}

Periodicity with respect to cell boundaries is introduced for both network geometry and boundary conditions.
This is achieved by using a method that was originally proposed in \cite{GraJir10} for two-dimensional analyses and then extended to three dimensions in \cite{AthWheGra17} for hydro-mechanical problems.
For every random point placed in the cell, 26 periodic image points in the adjacent cells are created.
The two dual tessellations are then performed for the points in the cell and the periodic image points.
In the resulting network, elements cross the boundaries of the cell.
In Figure~\ref{fig:periodic}, the periodic cell with two out of 26 adjacent cells is shown.

\begin{figure}[htbp!]
\begin{center}
  \begin{tabular}{cc}
    \includegraphics[width=5cm]{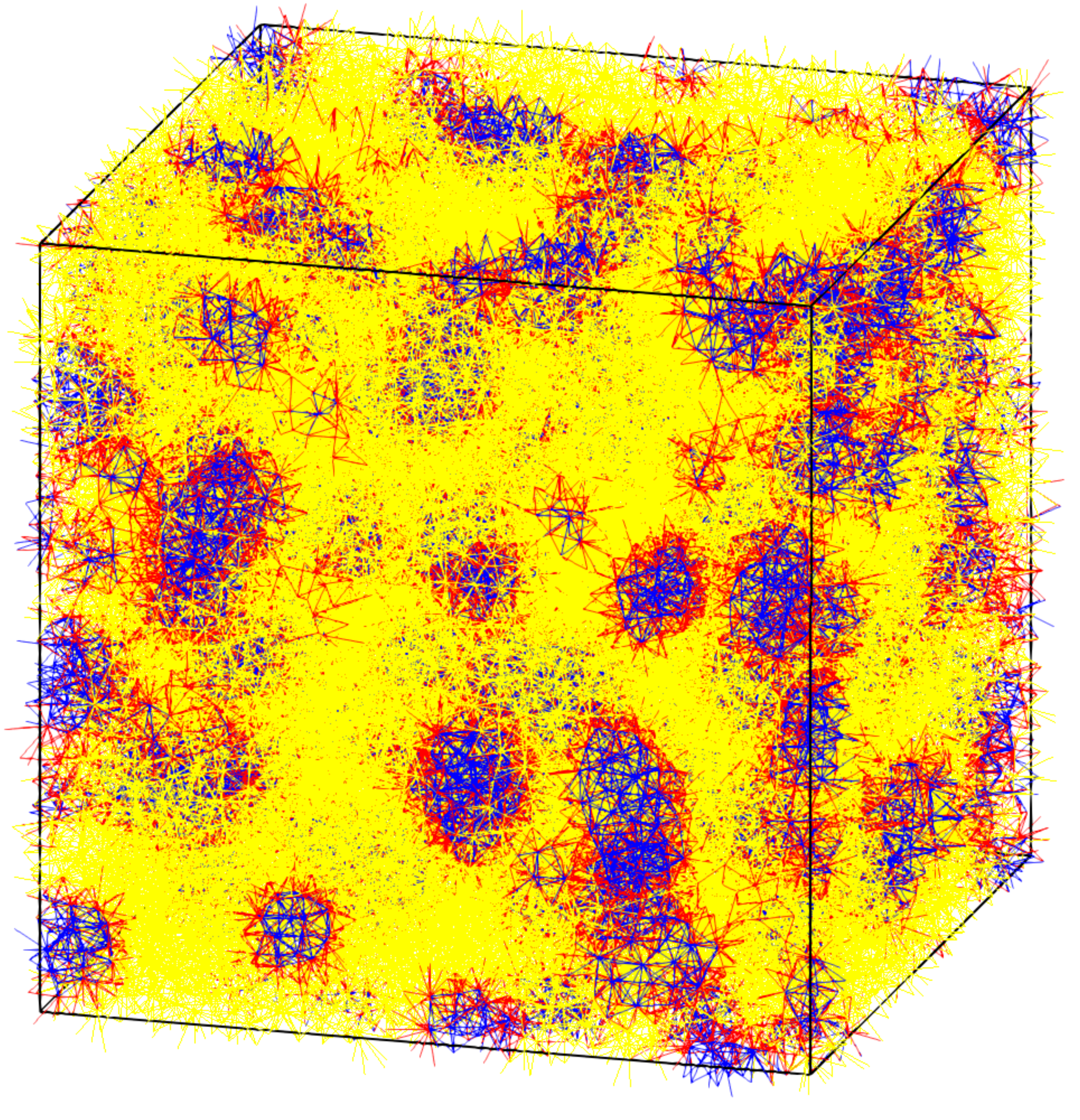} & \includegraphics[width=5cm]{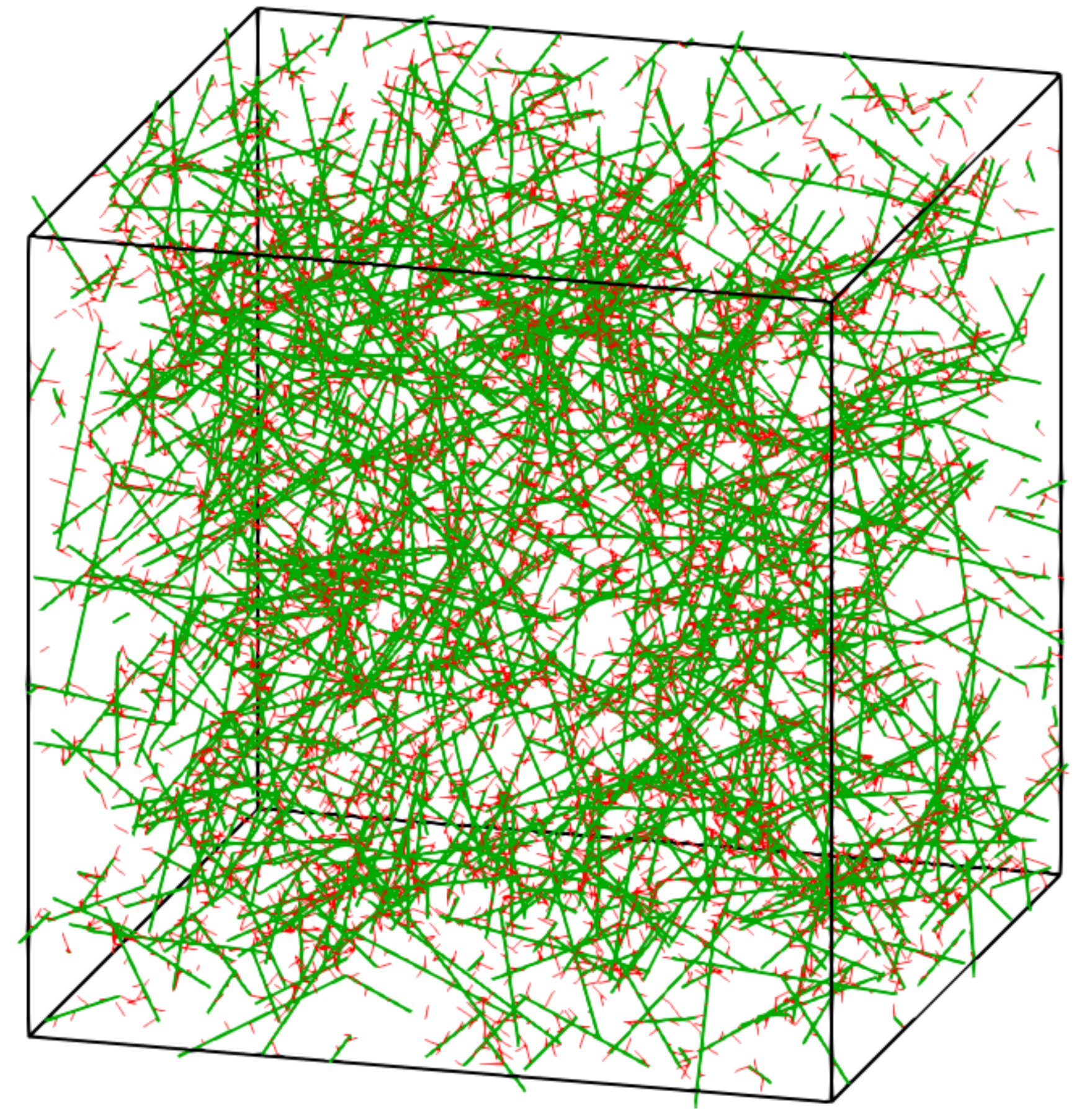}\\
    (a) & (b)
\end{tabular}
  \caption{Network model: (a) Network of discrete elements representing matrix (yellow), aggregates (blue) and ITZs (red). (b) Fibre frame elements (green) arranged independently of background network and links (red) connecting fibres to network nodes. Colours refer to online version.}
\label{fig:networkExamples}
\end{center}
\end{figure}

As an example, elements $I'-J$ and $I-J'$ cross the boundary of the cell.
These elements are used for computing the response of the periodic cell. However, only degrees of freedom (DOF) of nodes located inside the periodic cell are determined. For nodes outside the periodic cell ($I'$ and $J'$), which belong to elements crossing the boundary, the DOF are determined from those of the periodic image nodes inside the cell ($I$ and $J$, respectively) and six average strain components ($\varepsilon_{\rm xx}$,$\varepsilon_{\rm yy}$,$\varepsilon_{\rm zz}$,$\varepsilon_{\rm xy}$,$\varepsilon_{\rm yz}$,$\varepsilon_{\rm xz}$), which are applied to the cell. With these average strain components and the work conjugated stress components ($\sigma_{\rm xx}$,$\sigma_{\rm yy}$,$\sigma_{\rm zz}$,$\sigma_{\rm xy}$,$\sigma_{\rm yz}$,$\sigma_{\rm xz}$), the loading of the periodic cell is controlled. 
This approach has the advantage that localised fracture process zones can occur anywhere in the periodic cell along the direction of loading and are not strongly influenced by the boundaries of the cell. 
Analyses of boundary value problems without the use of periodic boundary conditions would normally require strengthening of the material close to the ends of the specimen to avoid fracture to occur at the boundaries. 
Furthermore, in alternative formulations of periodic boundary conditions, in which the elements close to the boundary are aligned so that the nodes are located on the boundary, the periodicity of the spatial arrangement of the network is not maintained.
A detailed description of the present periodic formulation can be found in \cite{GraJir10} and \cite{AthWheGra17}.
This approach is applied to both the background network and the frame and link elements. 
An example of the background network representing the three phases of matrix, aggregates and ITZ is shown in Figure~\ref{fig:networkExamples}a. Fibres with their corresponding link elements are shown in Figure~\ref{fig:networkExamples}b.

\subsection{Constitutive models} \label{sec:constitutive}

The constitutive response of the background network representing aggregates, matrix and ITZs are modelled by linear elasticity and damage mechanics.
For matrix and ITZ, a scalar damage model is used of the form
\begin{equation}
\boldsymbol{\sigma} = \left(1-\omega\right) \mathbf{D}_{\rm e} \boldsymbol{\varepsilon}
\end{equation}
where $\boldsymbol{\sigma}$ and $\boldsymbol{\varepsilon}$ are the stress vector and strain vector, respectively, $\mathbf{D}_{\rm e}$ is the elastic stiffness matrix and $\omega$ is the damage parameter ranging from 0 (undamaged) to 1 (fully damaged). For a detailed description of this constitutive model, see \cite{GraBol16}. 

\begin{figure}
  \begin{center}
\begin{tabular}{cc}
  \includegraphics[width=5cm]{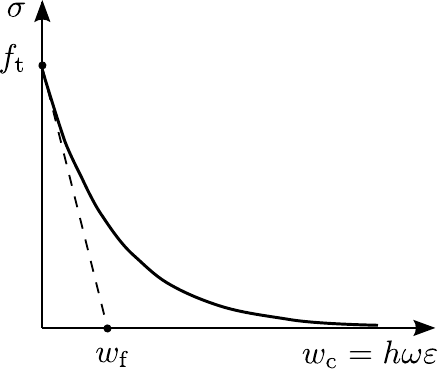} & \includegraphics[width=5cm]{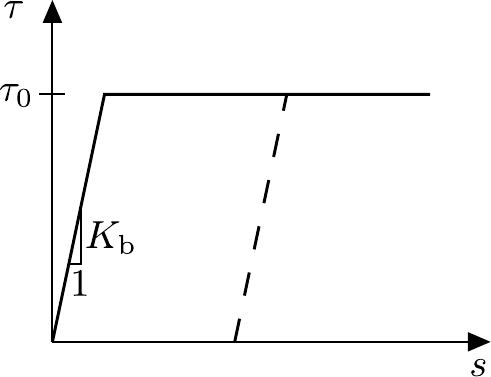}\\
  (a) & (b)
\end{tabular}
\caption{Constitutive models for a) softening in the matrix and b) bond-slip.}
\label{fig:constitutive}
  \end{center}
\end{figure}

By using the special network generation in Section~\ref{sec:network} and choosing the stiffness matrix $\mathbf{D}_{\rm e}$ so that the axial stiffness component is equal to the shear stiffness components, the stress and strain fields are elastically homogeneous and produce zero Poisson's ratio \citep{YipMohBol05}.
A global non-zero Poisson's ratio can be obtained by choosing lower shear than axial stiffness components. However, the elastic response is then no longer homogeneous as discussed in \citep{YipMohBol05}. This is a shortcoming of the present lattice approach, which can be overcome by techniques described in \citep{AsaAoyKim17}. The influence of the elastic Poisson's ratio on the results of the present analyses is very small, since the response is dominated by nonlinear processes.
The onset of damage is determined by an equivalent strain expression which gives an ellipsoidal strength envelope in the stress space with the shear and compressive strength being greater than the tensile strength as described in \cite{GraBol16}. The three input parameters for this strength envelope are the tensile strength $f_{\rm t}$, shear strength $f_{\rm q} = 2 f_{\rm t}$ and compressive strength $f_{\rm c} = 10 f_{\rm t}$.
The damage variable is determined from an exponential softening stress-crack opening curve ($\sigma$-$w_{\rm c}$) with tensile strength $f_{\rm t}$ and parameter $w_{\rm f}$, which controls the slope of the softening curve (Figure~\ref{fig:constitutive}). The area under the stress-crack opening curve is the fracture energy $G_{\rm F} = f_{\rm t} w_{\rm f}$. With this approach, the resulting load-displacement curves of tensile fracture simulations are independent of the element length, if the inelastic displacements localise in element length dependent zones. The dissipated energy rate $\dot{d}$ per unit cross-sectional area in the network element is computed as
\begin{equation} \label{dissipationMatrix}
  \dot{d} = h \dot{\omega} \dfrac{1}{2} \boldsymbol{\varepsilon} : \mathbf{D}_{\rm e} : \boldsymbol{\varepsilon}
\end{equation}
This dissipated energy is used in Section~\ref{sec:results} to present the fracture process zone.
Aggregates are assumed to be elastic. However, fracture in aggregates could be simulated in future studies with this approach, since aggregates are discretised by multiple network elements.

Fibres are modelled to be elastic with a Young's modulus $E_{\rm s}$. For the links between the fibres and the network model, an elasto-plastic model in the tangential direction of the fibre is used which is illustrated in Figure~\ref{fig:constitutive}b. Here, $\tau_0$ is the limit stress at which plastic slip $s$ occurs.
The stiffness $K_{\rm b}$ controls the elastic response of the link.
In the analyses, $K_{\rm b}$ is set to a large enough value (stated in Section~\ref{sec:results}) so that the results are not influenced significantly by it, but small enough so that no numerical problems are created.
The dissipated energy rate $\dot{d}$ per unit area of embedment for the constitutive model of the link element is
\begin{equation}\label{eq:dissFibres}
  \dot{d} = \left(s - s_{\rm p}\right) K_{\rm b} \dot{s}_{\rm p}
\end{equation}
Here, $\dot{s}_{\rm p}$ is the rate of plastic slip.

\subsection{Roughness evaluation} \label{sec:roughness}
\begin{figure}[htbp!]
\begin{center}
\includegraphics[width=8cm]{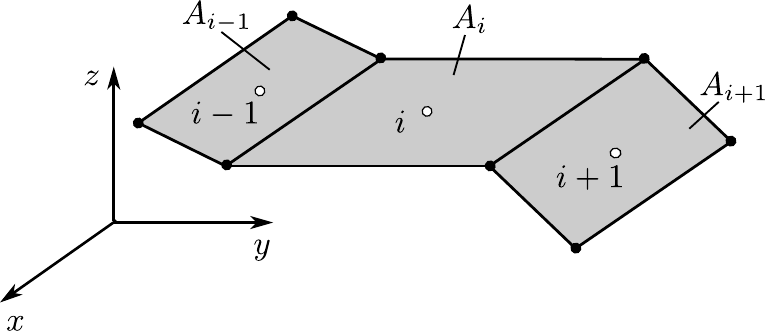} 
\caption{Evaluation of roughness from dissipated energy density of mid crosssections of network elements.}
\label{fig:tortuosity}
\end{center}
\end{figure}

The fracture processes are analysed by evaluating the evolution of spatial distribution of dissipated energy. 
For the present evaluation, both dissipation due to damage in the structural network elements, as well as dissipation due to plastic slip in the link elements are considered. 
To each element in which energy is dissipated, a crosssectional area with a centroid as shown in Figure~\ref{fig:tortuosity} is associated.
For the elements used for the background network, these are the mid-crosssections with the centroid $C$ shown in Figure~\ref{fig:voronoi}b.
For the link elements, the crosssectional area is $A_{\rm b}$ shown in Figure~\ref{fig:fibres}b and the centroid is the node of the frame element to which the link element is connected (node $H$ in Figure 5b).

Firstly, the mean of all heights of centroids of crosssections is calculated as
\begin{equation}
\bar z = \underset{i\rm=1}{\overset{N}{\sum}} w_{i} z_{i}
\end{equation}
Here, $z$ is measured in the direction of the applied tensile strain with the bottom of the cell used as the origin.
Furthermore, $w_{i}$ are the weights of the individual crosssections, which are calculated as 
\begin{equation}
w_{i} = \dfrac{ A_{i} \Delta d_{i}}{\underset{k \rm =1}{\overset{N}{\sum}} A_{k} \Delta d_{k} } \\
\end{equation}
where $A_{i}$ and $\Delta d_{i}$ are the area and increment of dissipation per unit area, respectively, of the facet $i$. 
Then, the standard deviation $\Delta h$ is calculated as 
\begin{equation}\label{eq:DeltaH}
\Delta h = \sqrt{ \underset{i \rm =1}{\overset{N}{\sum}} w_{i} (z_{i}-\bar z)^2}
\end{equation}
This standard deviation is a measure related to the width of the fracture process zone, which takes into account the spatial arrangement and intensity of the dissipation events.
It is smaller than the total width of the fracture process zone, which is simply defined as the zone in which energy is dissipated, but does not provide information about the intensity of these events.
For a localised crack surface with equal energy dissipation in all elements whose crosssections form this surface, the measure used is equal to the standard deviation of the roughness distribution of the crack surface, which can be determined experimentally as described in \cite{XenGreMorGra15}.
Because of this geometrical link to the fracture surface, the method is called here roughness evaluation.
Nevertheless, for energy dissipation in overlapping zones and fibres, it would not be possible to determine the value of $\Delta h$ experimentally by means of evaluation of the roughness of the surface alone.

\section{Analyses} \label{sec:analyses}
\begin{figure}[htbp!]
\begin{center}
\includegraphics[width=5cm]{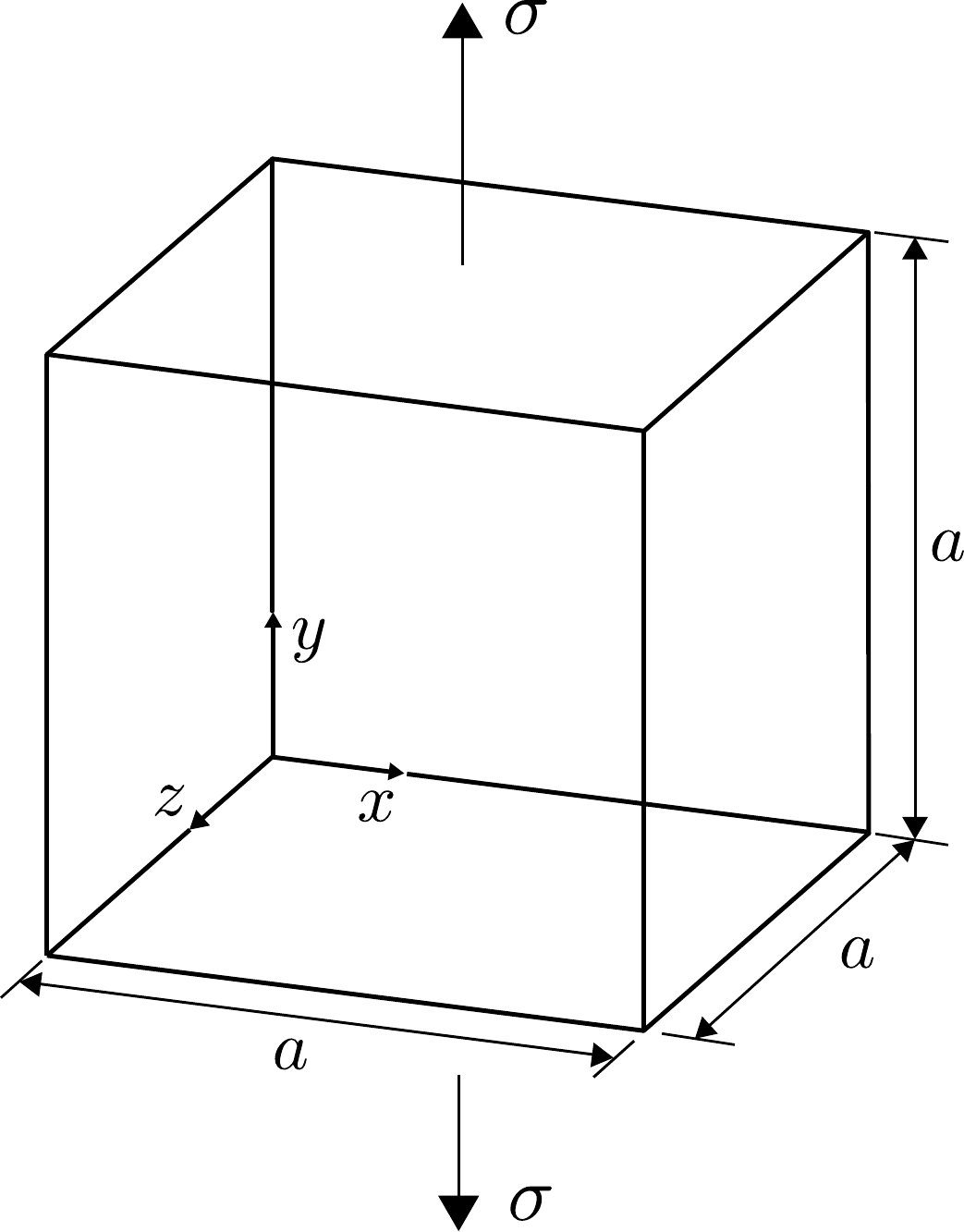} 
\end{center}
\caption{Setup for direct tension analysis with the periodic cell.}
\label{fig:setup}
\end{figure}

The network modelling approach described in Section~\ref{sec:method} is applied to analyse fracture in cubic periodic cells of an edge length of $100$~mm subjected to direct tension as shown in Figure~\ref{fig:setup}.
For this setup, the average strain in the axial $y$-direction ($\varepsilon_{\rm yy}$) is monotonically increased, which results in a reactive stress component in the $y$-direction ($\sigma_{\rm yy}$), which in the presentation of the results is called $\sigma$.
All other average stress components ($\sigma_{\rm xx}$, $\sigma_{\rm zz}$, $\sigma_{\rm xy}$, $\sigma_{\rm yz}$ and $\sigma_{\rm xz}$) are kept equal to zero.
The analyses are performed quasi-statically with an incremental-iterative approach (see e.g. \cite{DebCrisRem12}). The iterative part is based on a modified Newton method using the secant stiffness for the damage model for matrix and ITZ, and the elastic stiffness for the elasto-plastic model for the links between fibres and background network.

Four groups of analyses are carried out. For each group, ten random generations of background networks and meso-structures are performed.
The network is generated with a minimum distance $d_{\rm min}=3$~mm between the randomly placed points.
The first group of analyses consists of a network representing matrix without any meso-scale features explicitly incorporated.
In the second group of analyses, the network of elements represented matrix, aggregates and ITZs.
For these analyses, the volume fraction of aggregates generated with the techniques described in Section~\ref{sec:meso} is $\rho_{\rm a} = 0.8$ with a maximum and minimum sieve size $d_{\rm a,max} = 16$~mm and $d_{\rm a,min} = 8$~mm, respectively.
In the third group of analyses, fibres with a length $l_{\rm f} =3$~cm, a diameter $d_{\rm f}=0.75$~mm and a fibre volume fraction of $\rho_{\rm f} = 0.01$ are used.
Finally, the fourth group consists of combinations of aggregates and fibres with the same input as for the analyses with only one phase.
The input parameters for the different phases of the background network are shown in Table~\ref{tab:input}.
These input values are in the typical range of values used for meso-scale analyses of concrete in the literature \citep{GraGreSol12}, where it was shown that they provide good agreement with experimental results.
For fibres, a modulus of $E_{\rm f}=200$~GPa is used.
The elastic stiffness and limit stress of the link elements is set to $K_{\rm b} = 3000$~GPa and $\tau_{\rm 0} = 4$~MPa, respectively.

\begin{table}
  \begin{center}
    \caption{Input values for the background network. The modulus $E$ of ITZ is determined as harmonic mean of moduli of matrix and particle.}
    \label{tab:input}
    \begin{tabular}{cccc}
      \hline
Phase & $E$~[GPa] & $f_{\rm t}$~[MPa] & $G_{\rm F}$~[J/m$^2$]\\
\hline
Matrix & 30 & 3 & 100\\
Particle & 90 & - & -\\
ITZ & 45 & 1.5 & 50\\
\hline\\
\end{tabular}
\end{center}
\end{table}

\section{Results} \label{sec:results}

The results of the direct tension analyses of the four groups of material setups are shown in the form of stress-displacement curves, spatial patterns of dissipated energy and roughness-displacement curves.
The displacement is determined as the average strain multiplied by the cell length $a$ (Figure~\ref{fig:setup}).
For the stress-displacement and roughness-displacement curves, the mean of the quantities of random analyses are shown.

\begin{figure}[htbp!]
\begin{center}
 \includegraphics[width=12cm]{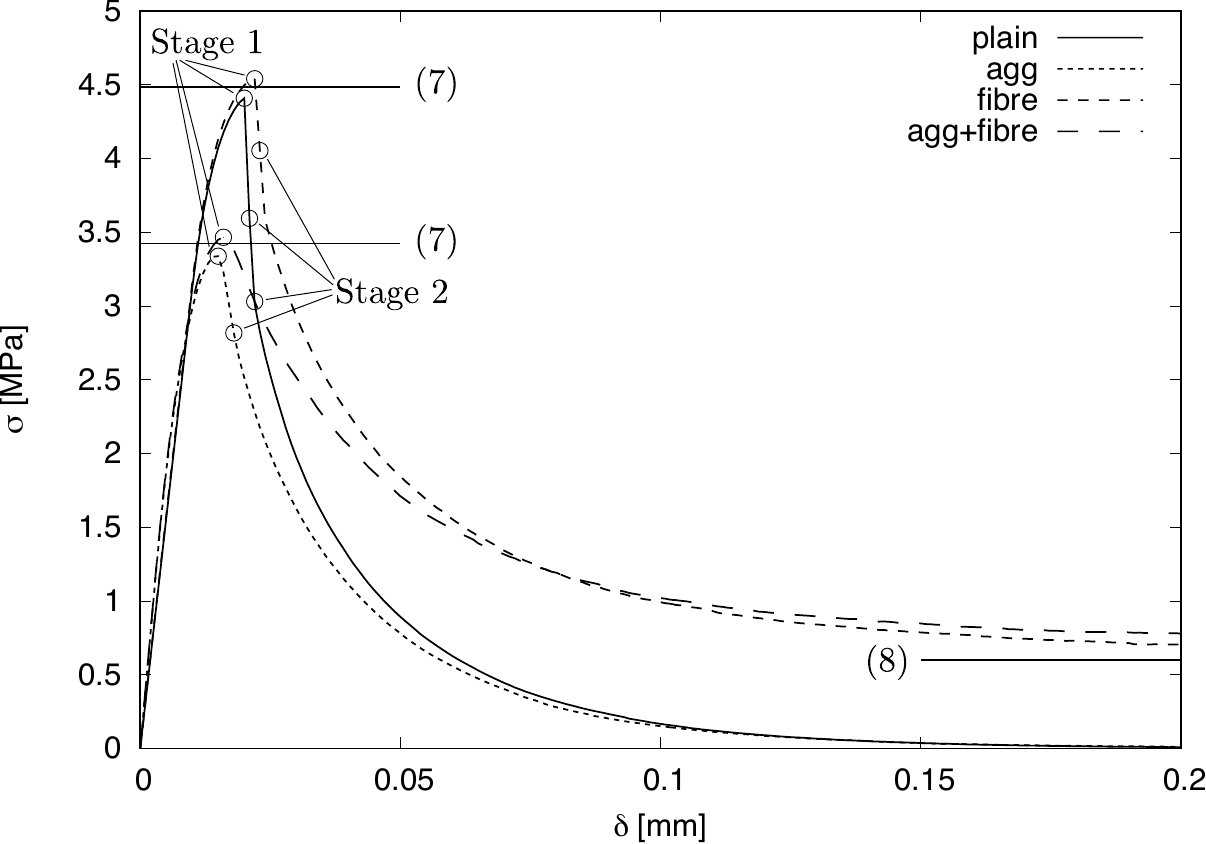}\\
 \caption{Meso-scale analysis: Mean stress versus displacement for four groups of material setups (plain, aggregates, fibres and aggregates+fibres). The symbols refer to stages for which the crack patterns are shown in Figures~\ref{fig:cracksStage1}~and~\ref{fig:cracksStage2}. Furthermore, the lines refer to empirical estimates in (\ref{eq:empiricalPeak}) and (\ref{eq:empiricalBridge}).}
 \label{fig:ld}
\end{center}
\end{figure}

The mean stress-displacement curves for four groups of material setup are shown in Figure~\ref{fig:ld}.
For the plain configuration with matrix material only, the stress-displacement curve showed the typical response of quasi-brittle materials subjected to direct tension.
In the pre-peak, the response is linear elastic in the first part and then exhibits small non-linearities just before the peak.
The post-peak regime shows steep softening, which then flattens with the average stress approaching zero.
The peak stress is greater than the input tensile strength, because the stress in the network elements consists of combinations of axial and shear components.
With the ellipsoidal strength envelope used, the combined normal and shear stress components result in a greater strength than a pure tensile stress component.
The addition of aggregates strongly reduces the peak stress because of the weak ITZs between aggregates and matrix.
Furthermore, the initial stiffness is slightly increased due to the greater stiffness of the aggregates.
If instead of aggregates only fibres are added to the matrix, the peak stress is only slightly increased compared to the plain peak stress.
However, the tail of the stress-displacement curve is strongly influenced by the presence of the fibres with a significant bridging stress present after the initial softening.
For combinations of aggregates and fibres, the fibres cause again a small increase of the peak stress compared to the aggregate only case and result in a similar bridging stress at the ultimate displacement applied in the analyses.

For the analyses involving fibres, the peak and bridging stresses are compared to empirical estimates reported in \cite{Naa87}.
For the peak values of the stress of the analyses with fibres, the peak stress is
\begin{equation} \label{eq:empiricalPeak}
\sigma_{\rm cc} = \sigma_{\rm mu} \left(1-\rho_{\rm f}\right) + \alpha_1 \alpha_2 \tau_{\rm 0} \rho_{\rm f} \dfrac{l_{\rm f}}{d_{\rm f}}
\end{equation}
Here, $\alpha_{\rm 1}$ and $\alpha_{\rm 2}$ are factors taking into account the fibre orientation and fraction of bond strength mobilised, respectively.
Furthermore, $\sigma_{\rm mu}$ is the peak stress of the material without fibres.
The stress after cracking is estimated as
\begin{equation}  \label{eq:empiricalBridge}
  \sigma_{\rm pc} = 4 \lambda_1 \lambda_2 \tau_{\rm 0} \rho_{\rm f} \dfrac{l_{\rm f}}{d_{\rm f}}
\end{equation}
where $\lambda_{1}$ and $\lambda_{2}$ are factors for average pullout length and postcracking orientation efficiency, respectively.
These expressions are compared to the numerical results in Figure~\ref{fig:ld} using $\alpha_1 = 0.5$, $\alpha_2 =  0.2$, $\lambda_1 = 0.25$ and $\lambda_2 = 0.5$, which are typical values for the type of fibres used.
It should be noted that (\ref{eq:empiricalPeak}) only predicts the increase of strength due to the presence of fibres, which is very small. The values for $\sigma_{\rm mu}$ in (\ref{eq:empiricalPeak}) are obtained for the corresponding analyses without fibres.  

\begin{figure}[htbp!]
  \begin{center}
    \begin{tabular}{cccc}
      \includegraphics[width=3.5cm]{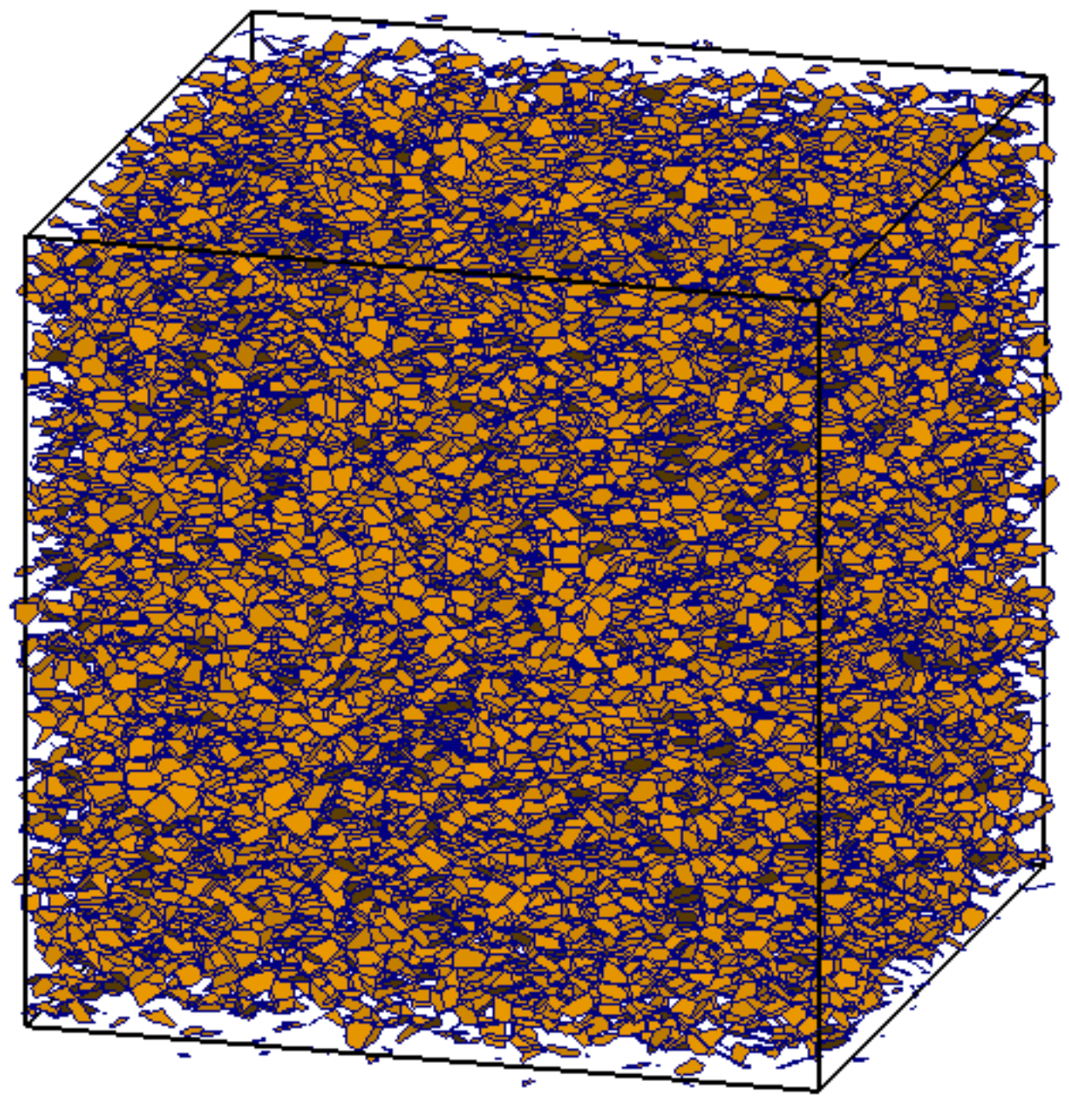} &
      \includegraphics[width=3.5cm]{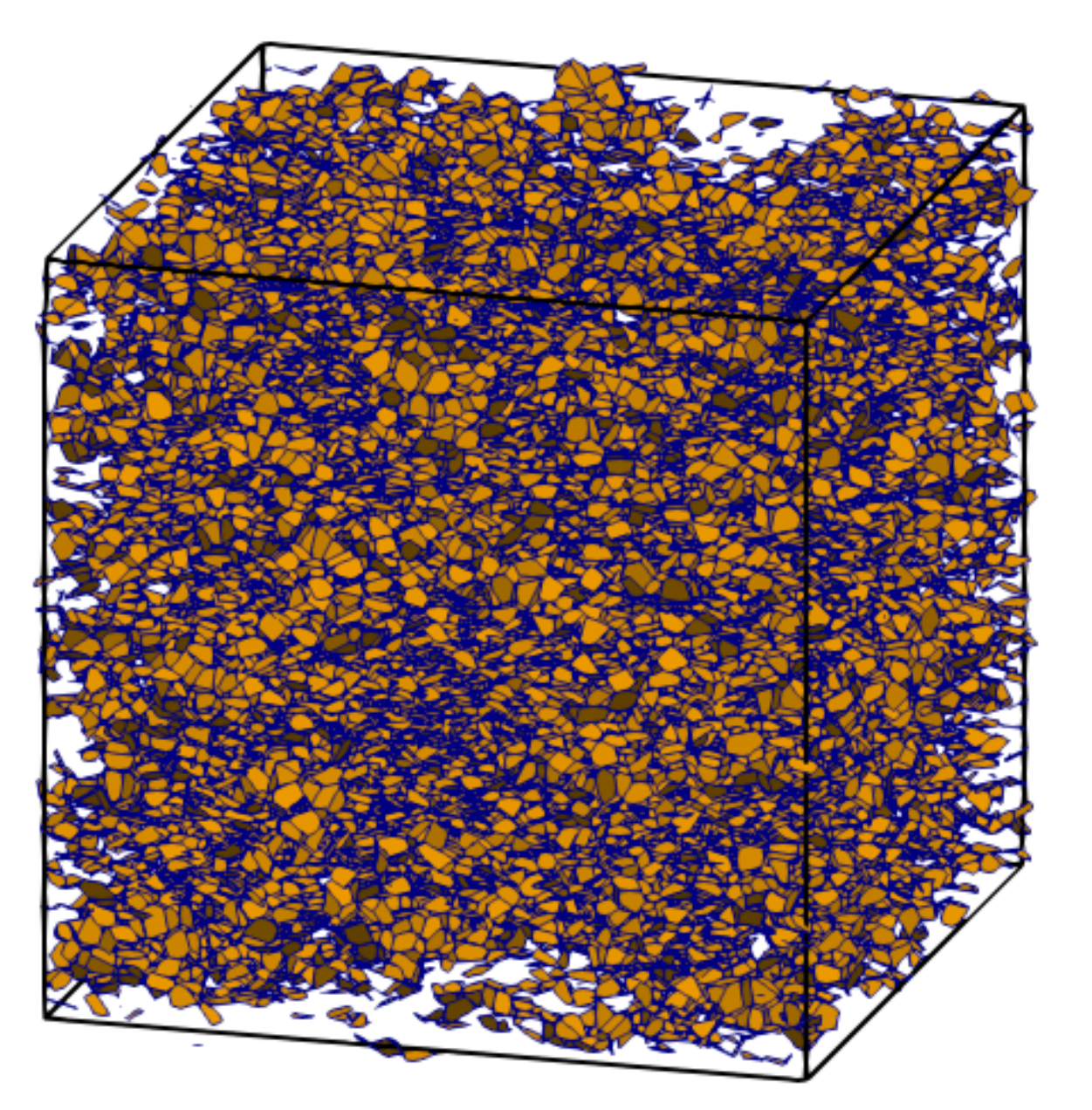} &
      \includegraphics[width=3.5cm]{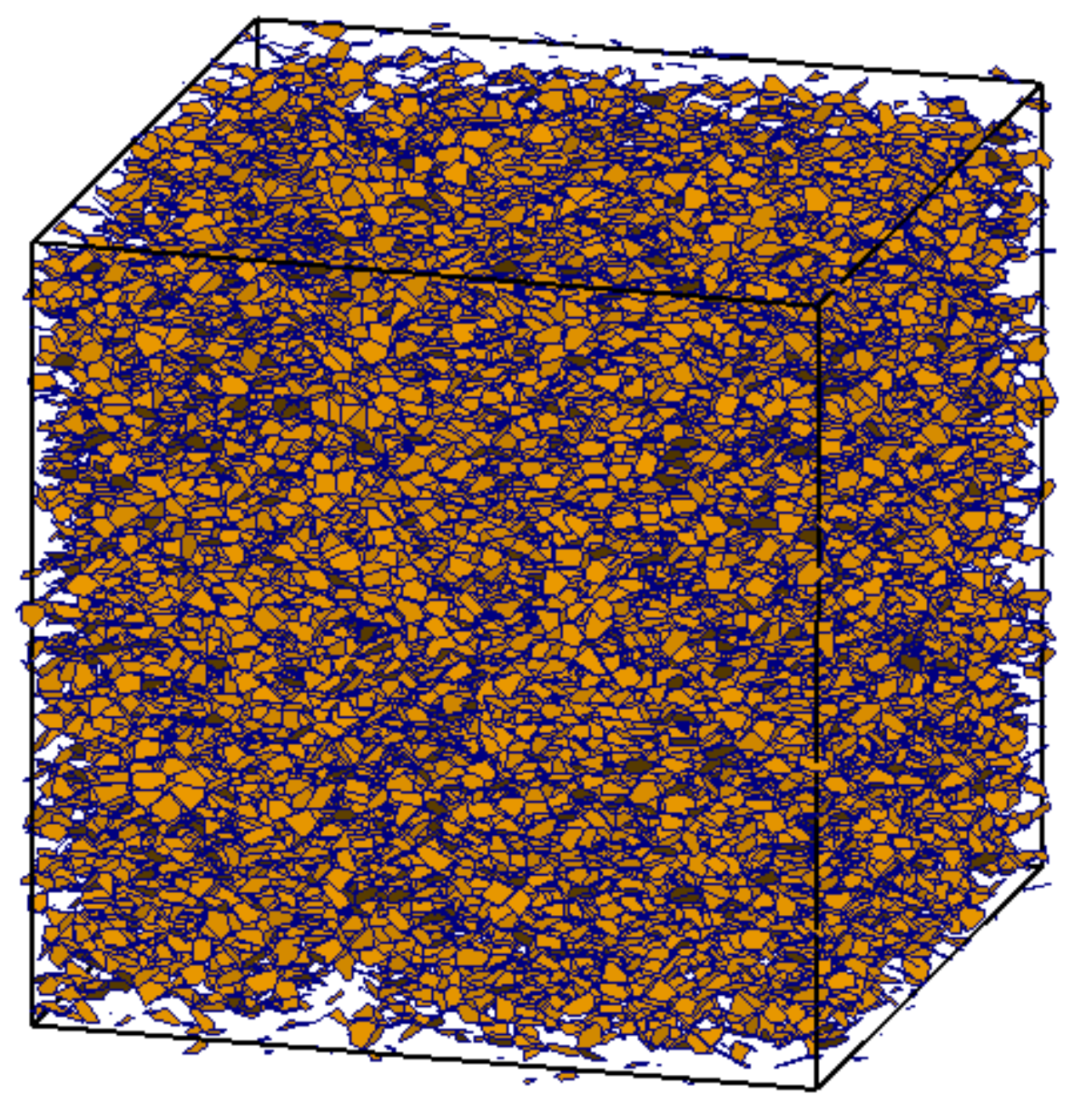} &
      \includegraphics[width=3.5cm]{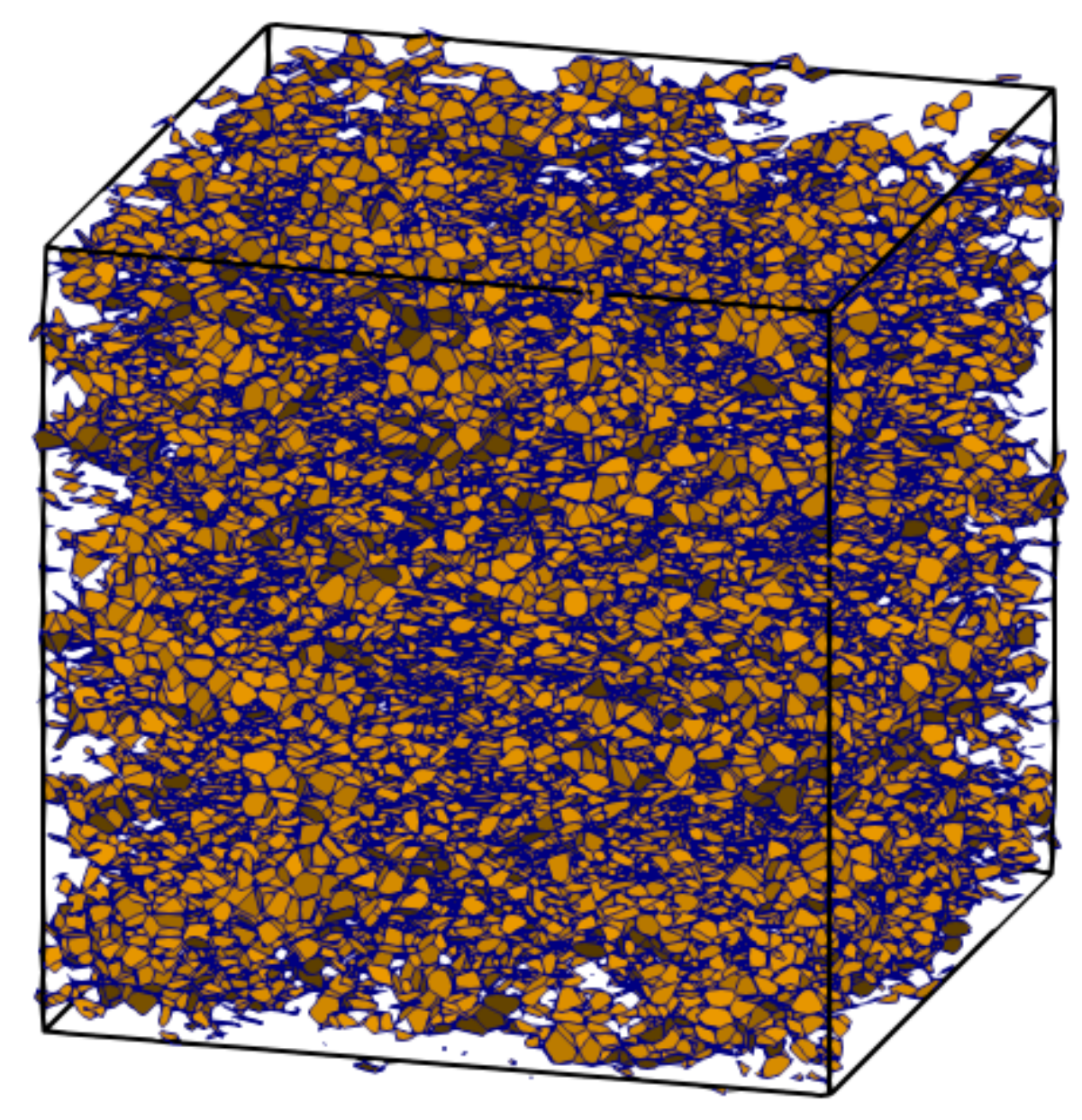} \\
      plain & agg & fibre & agg+fibre \\
    \end{tabular}
 \caption{Meso-scale analyses: Crack patterns of direct tension analysis at stage~1 marked in Figure~\ref{fig:ld}. Orange (online version) polygons refer to mid crosssections in which damage increases at this stage of the analysis.}
 \label{fig:cracksStage1}
  \end{center}
\end{figure}

\begin{figure}[htbp!]
  \begin{center}
    \begin{tabular}{cccc}
      \includegraphics[width=3.5cm]{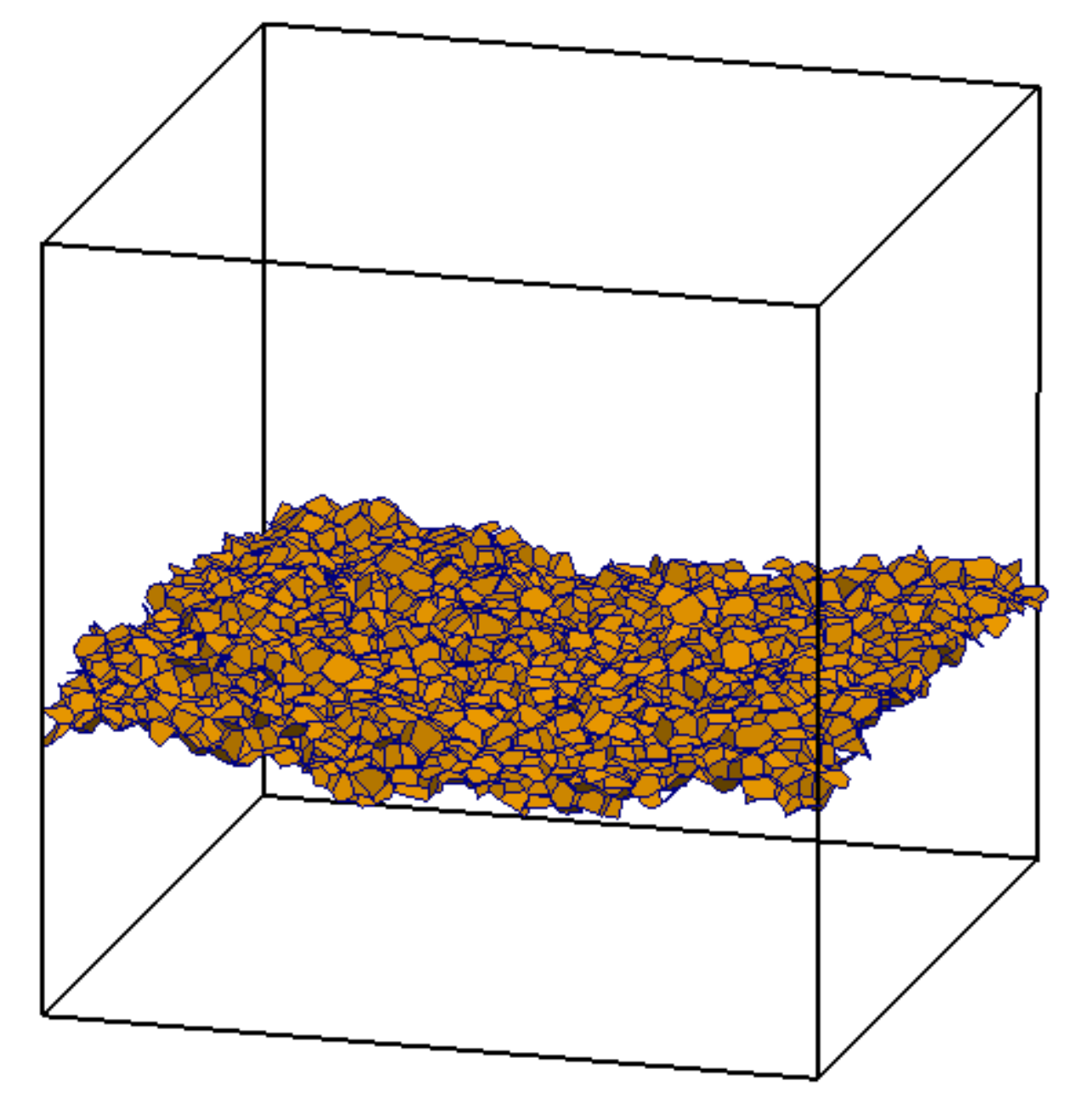} &
      \includegraphics[width=3.5cm]{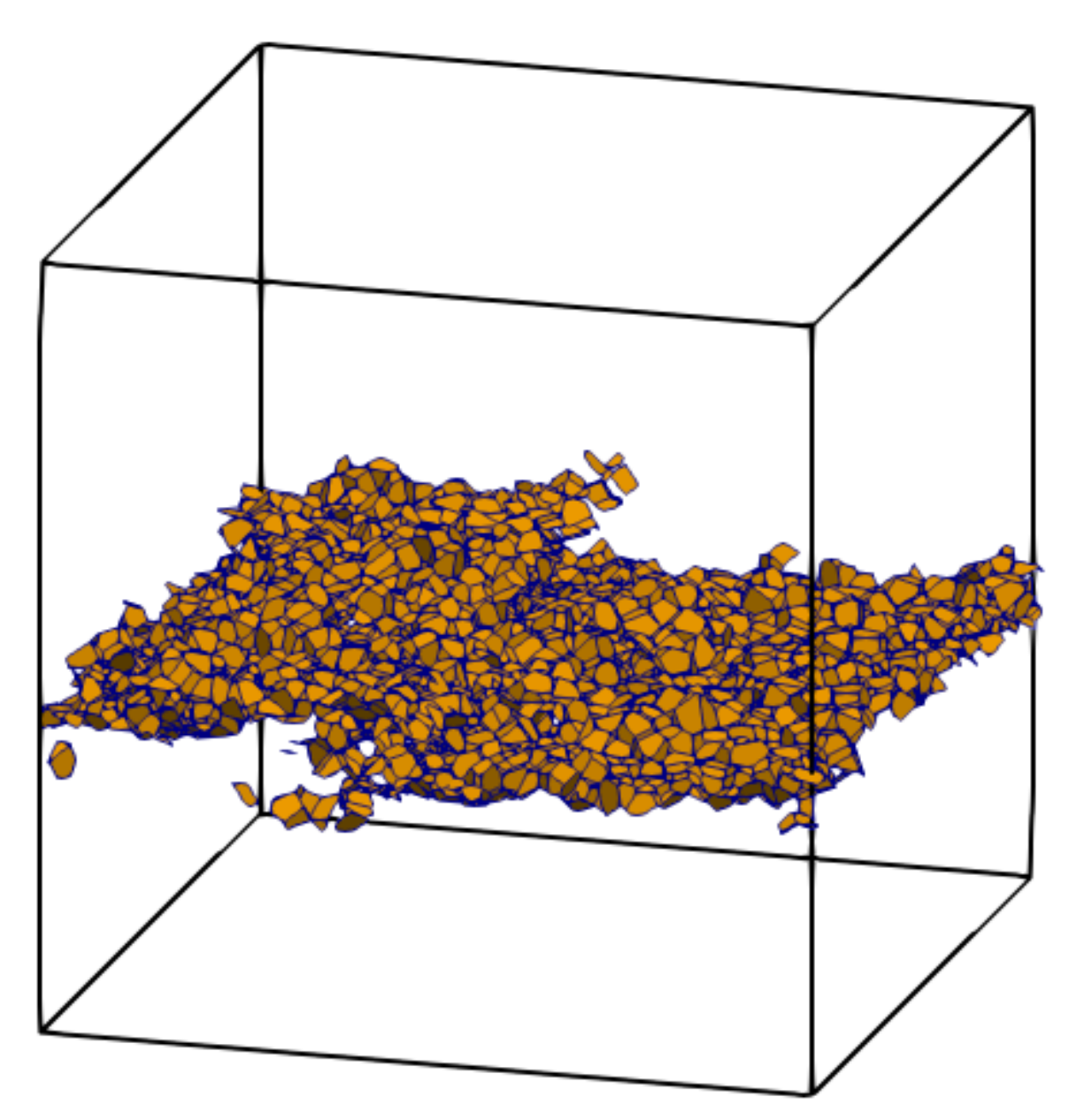} &
      \includegraphics[width=3.5cm]{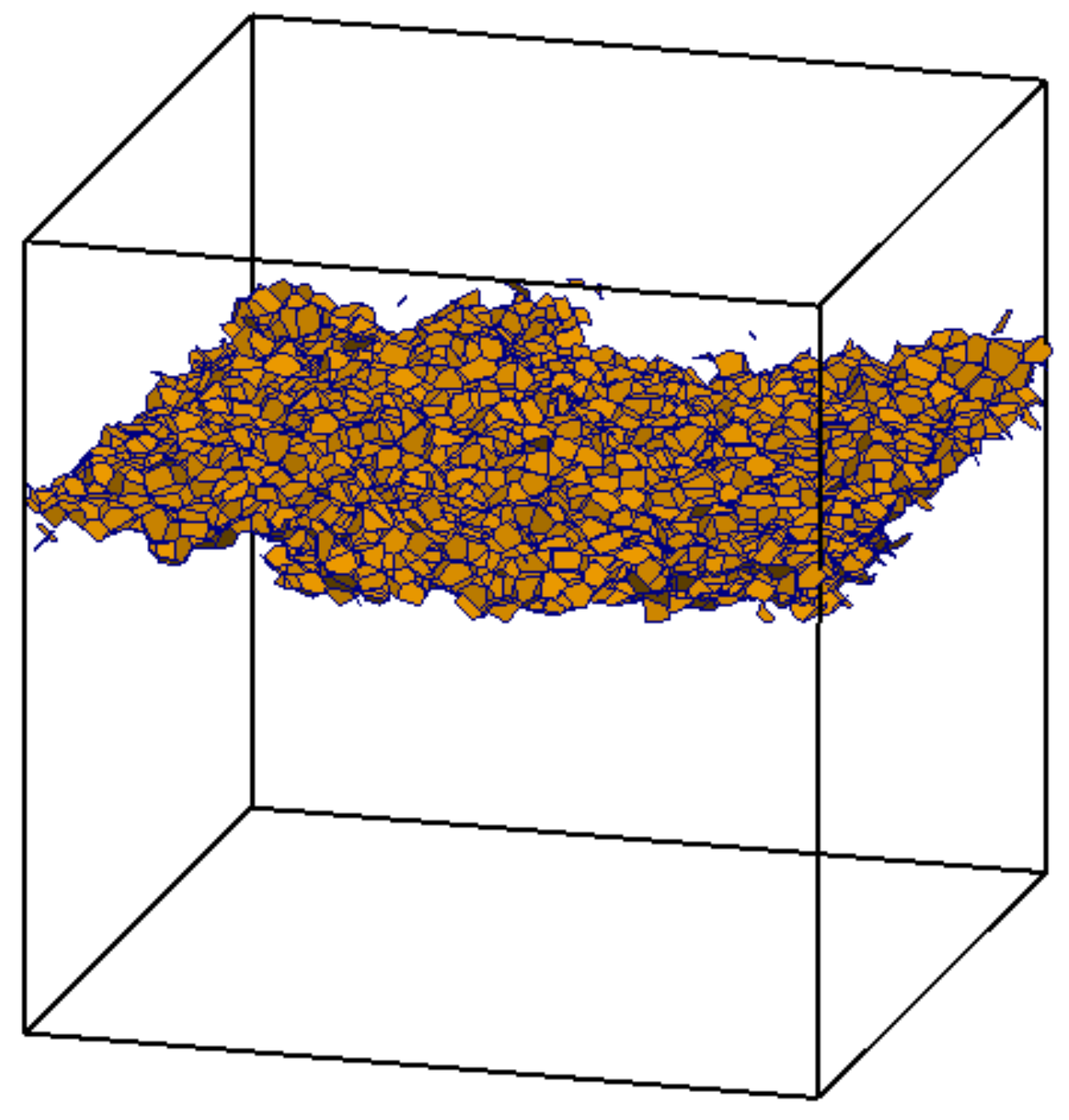} &
      \includegraphics[width=3.5cm]{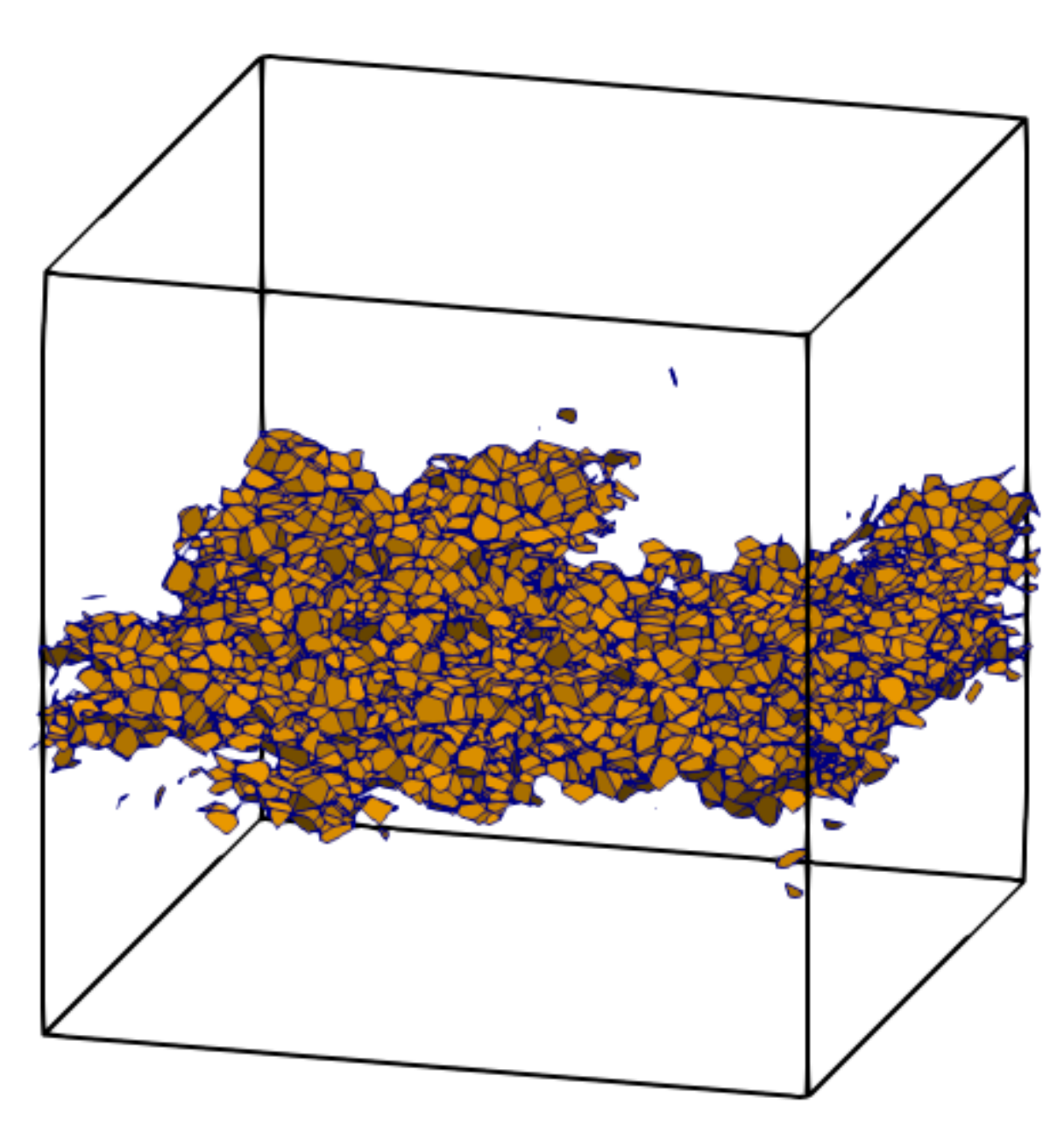}\\
      plain & agg & fibre & agg+fibre \\
    \end{tabular}
 \caption{Meso-scale analyses: Crack patterns of direct tension analyses at stage 2 marked in Figure~\ref{fig:ld}. Orange (online version) polygons refer to mid crosssections in which damage increases at this stage of the analysis.}
  \label{fig:cracksStage2}
  \end{center}
\end{figure}

All the gobal stress-displacement curves in Figure~\ref{fig:ld} exhibit softening which is usually accompanied by localisation of displacements.
Detailed information about the localisation process is studied in the form of spatial distribution of mid-crosssections at which energy dissipation occurs.
The dissipation patterns for the four groups of analyses are shown in Figures~\ref{fig:cracksStage1}~and~\ref{fig:cracksStage2} for stages at peak and in the post-peak, respectively, for one random analysis.

The corresponding stages are marked in Figure~\ref{fig:ld}. At stage~1 at peak, the dissipation rate is distributed in the entire specimen (Figure~\ref{fig:cracksStage1}).
For plain and fibre analyses, the dissipated energy is distributed uniformly.
For analyses involving aggregates, the distribution is more heterogeneous, because at the position of the elastic aggregates no dissipation occurs.
At stage~2 in the softening regime, the rate of dissipation is strongly localised (Figure~\ref{fig:cracksStage2}).
The y-position of the localised region of rate of dissipation differs from the analysis to analysis because of the periodic cell used.
For all groups, the localised zone is rough.
For the plain analyses, this is due to the irregular background network used.
For the other groups, the roughness of the zone of dissipated energy is also influenced by the heterogeneity in the form of aggregates and fibres.
For instance, the spatial distribution of energy for the aggregate analyses in Figure~\ref{fig:cracksStage2} appears to be wider than for the plain case.
These plots of dissipation rate are from only one random analysis of each group.
Also, all mid-crosssections at which energy is dissipated at this stage of the analysis are shown without discriminating between the amount of energy that is dissipated at the crosssections.

\begin{figure}[htbp!]
\begin{center}
 \includegraphics[width=12cm]{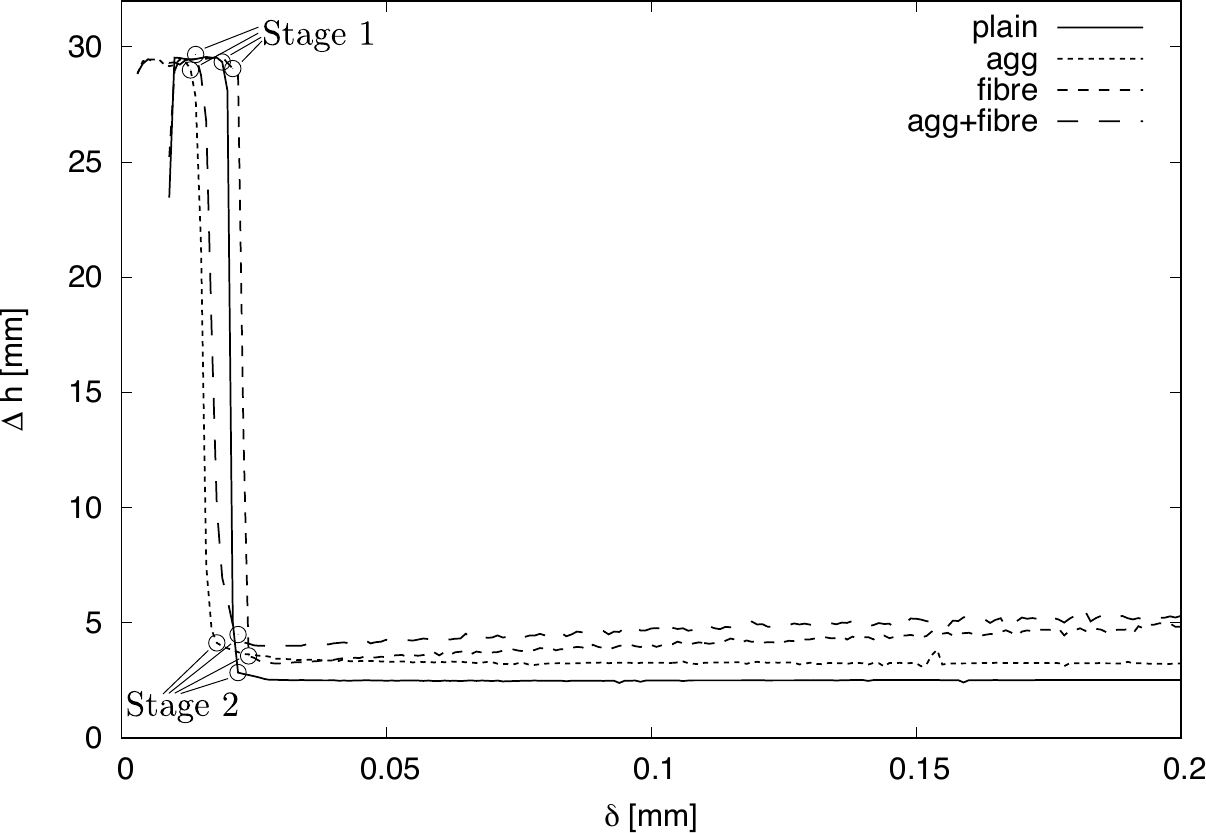}\\
 \caption{Meso-scale analysis: Measure of width of fracture process zone $\Delta h$ versus displacement $\delta$ for random analyses with aggregates, and with aggregates and fibres. The symbols refer to stages for which the crack patterns are shown in Figures~\ref{fig:cracksStage1}~and~\ref{fig:cracksStage2}.}
\label{fig:roughness}
\end{center}
\end{figure}

For a quantitative representation of the evolution of the zone of rate of dissipated energy, the roughness measure described in Section~\ref{sec:roughness} is used.
The mean of the measure of the width of the fracture process zone $\Delta h$ in (\ref{eq:DeltaH}) versus displacement is shown in Figure~\ref{fig:roughness}.
The symbols in the figure refer to the two stages at which the crack patterns are shown in Figures~\ref{fig:cracksStage1}~and~\ref{fig:cracksStage2}.
The overall roughness evolutions for the four groups of analyses are overall very similar.
At the start of the analysis, no energy is dissipated, so that $\Delta h$ is not defined. For the uniformly distributed cracking in pre-peak regime, $\Delta h$ is approximately equal to $30$~mm.
This value agrees well with the theoretical value for the standard deviation of a uniform distribution over the cell size, i.e. the interval from $0$~to~$100$~mm, which is $100/\sqrt{12} = 28.9$~mm.
At the start of the post-peak regime, the width of the fracture process zone drops down to values less than $5$~mm for all groups of analyses.
This drop occurred in the initial part of the softening regime at a stage at which little energy had been dissipated.

\begin{figure}[htbp!]
\begin{center}
 \includegraphics[width=12cm]{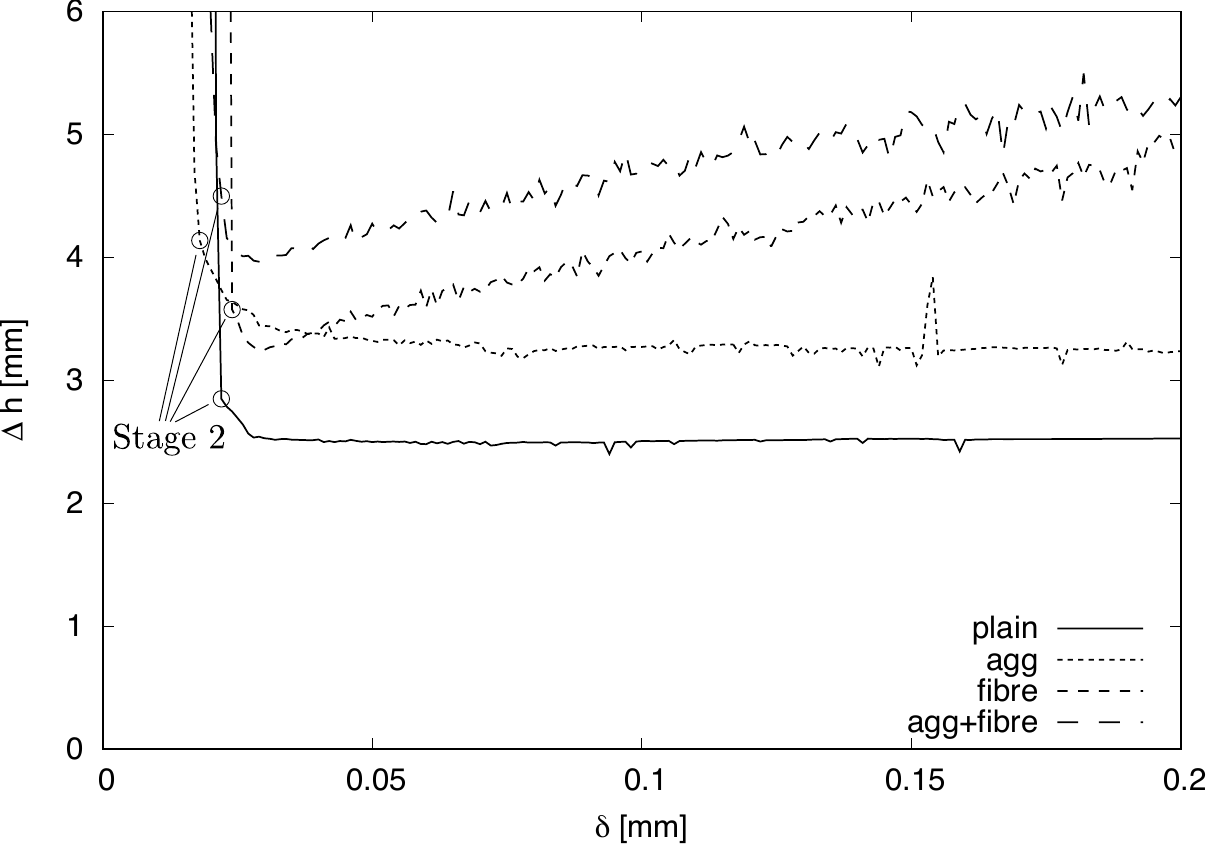}\\
 \caption{Meso-scale analysis: Stress versus displacement analyses with aggregates and aggregates with fibres. The symbols refer to stages for which the crack patterns are shown in Figures~\ref{fig:cracksStage1}~and~\ref{fig:cracksStage2}.}
 \label{fig:roughnessDetail}
\end{center}
\end{figure}

A detail of the evolution of $\Delta h$ after the drop is shown in Figure~\ref{fig:roughnessDetail}.
The roughness $\Delta h$ is the smallest for the analyses with only the matrix material.
After the abrupt drop, $\Delta h$ remains almost constant.
Adding aggregates results in an increase of the roughness compared to the plain case.
Again, the value remains constant after the drop. 
If, instead of aggregates, fibres are added to the background lattice, the roughness is again greater than for the plain case.
However, roughness is not constant with increasing displacement.
Instead, it increases with increasing displacement.
The same trend is observed if aggregates and fibres are combined.
This increase is due to the energy dissipated by the slip between fibre and matrix defined in (\ref{eq:dissFibres}).
Before the abrupt drop, there is no energy dissipation due to fibre slip.
Only once the crack has formed, the slip between fibres and matrix starts.
In the present approach, fibre pull out is not modelled, which means that the embedded length of fibres does not change.
Consequently, it is expected that for the analyses involving fibres, $\Delta h$ would reach a constant value once all fibres crossing the localised zone of displacements are significantly stressed so that they dissipate energy along their short embedded length, and the damage in the matrix so high that the energy dissipation in the matrix is insignificant.
If fibre pullout would be taken into account as well, the dissipated energy should eventually reduce to zero once all fibres are pulled out.
For a fibre length of $3$~cm as used in this study, this point would be reached when a displacement of $1.5$~cm is applied to the specimen, which is $100$ times higher than the maximum displacement considered here.

\begin{figure}[htbp!]
\begin{center}
 \includegraphics[width=12cm]{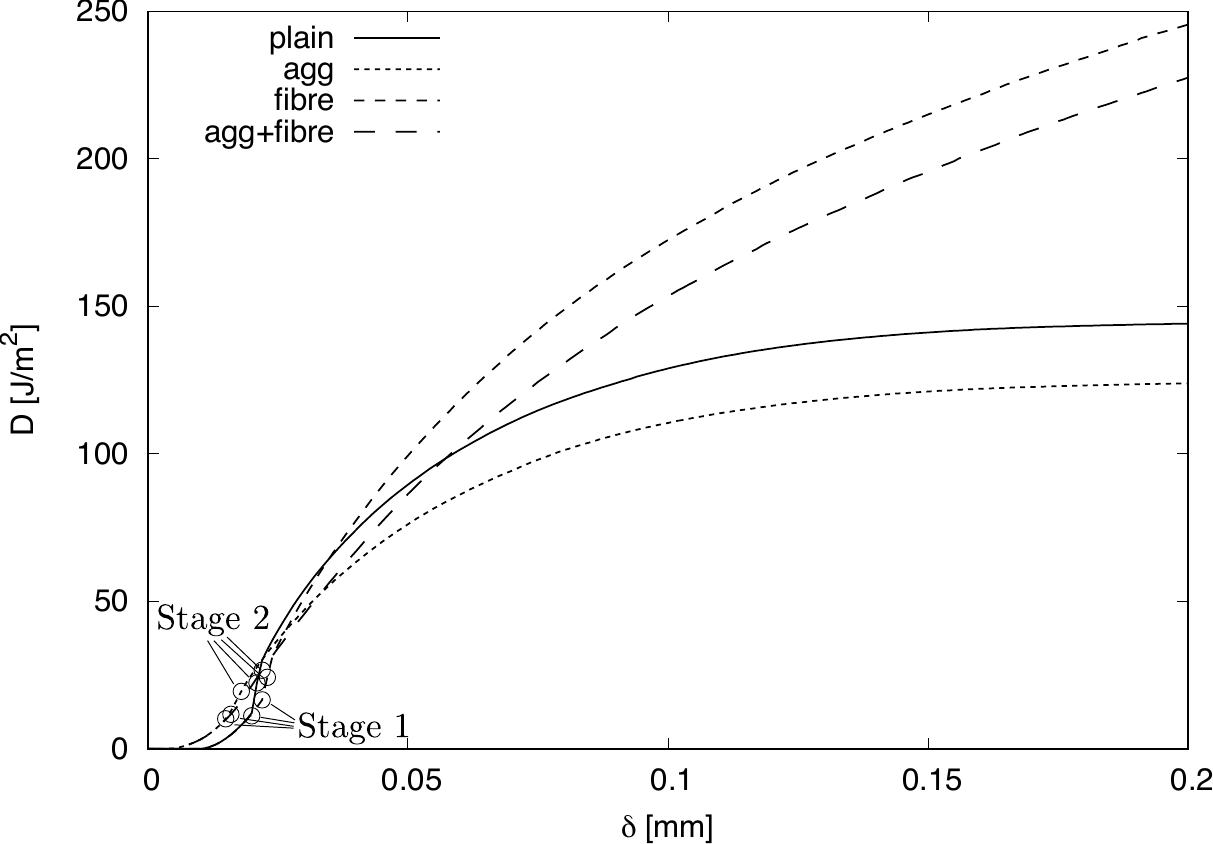}\\
 \caption{Meso-scale analysis: Dissipated energy $D$ versus displacement $\delta$ for the four groups of analyses (plain, aggregates, fibres and aggregates+fibres). The symbols refer to stages for which the crack patterns are shown in Figures~\ref{fig:cracksStage1}~and~\ref{fig:cracksStage2}.}
\label{fig:energy}
\end{center}
\end{figure}

The evolution of dissipated energy for the four groups of analyses is shown in Figure~\ref{fig:energy}.
The symbols refer to the two stages at which the dissipation patterns are shown in Figures~\ref{fig:cracksStage1}~and~\ref{fig:cracksStage2}.
Here, stage~1 marks the peak of the stress-displacement curves shown in Figure~\ref{fig:ld}.
For all analyses, the dissipation in the pre-peak regime is very small.
For plain and aggregates only cases, the majority of dissipation occurs in the first part of the post-peak regime and then approaches a constant value.
For the analyses with fibres, the initial dissipation in the very first part of the post-peak regime is slightly less than for the analysis without fibres.
However, this difference is very small. 
In the later stage of the post-peak regime, the fibres contribute significantly to the dissipation, so that the overall dissipation of the analyses with fibres is much greater than for aggregates only.
Only fibres, which cross the localised zone shown in Figure~\ref{fig:cracksStage2}, are stretched sufficiently to contribute to the dissipated energy.

\begin{figure}[htbp!]
\begin{center}
 \includegraphics[width=12cm]{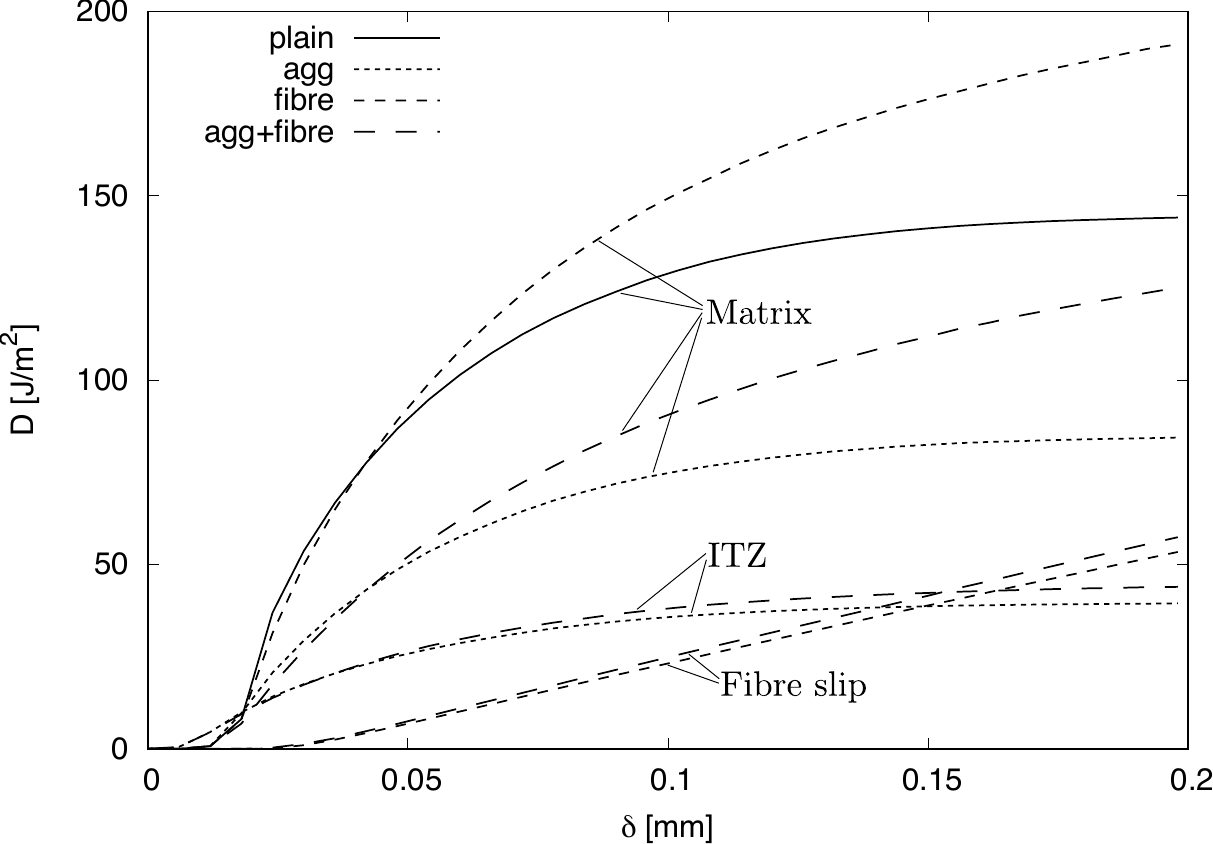}\\
 \caption{Meso-scale analysis: Dissipated energy $D$ versus displacement $\delta$ for the four groups of analyses (plain, aggregates, fibres and aggregates+fibres) in the three phases of material in which energy is dissipated (matrix, ITZ, link be. The symbols refer to stages for which the crack patterns are shown in Figures~\ref{fig:cracksStage1}~and~\ref{fig:cracksStage2}.}
 \label{fig:energyComponents}
\end{center}
\end{figure}

The interplay of energy dissipation in the different phases (matrix, ITZ and slip between fibres and matrix) is illustrated for the four groups of analyses in Figure~\ref{fig:energyComponents}.
From this figure, it can be seen that fibres only contribute to the dissipation in the post-peak regime of the stress-displacement curve in Figure~\ref{fig:ld}.
Furthermore, the matrix material dissipates more energy if fibres are present, which is most likely due to the generation of multiaxial stress states in the material.
The dissipation within the ITZs is not affected by the presence of the fibres, since the majority of energy dissipation in the ITZs occurs early in the fracture process before the fibres are activated.
Furthermore, fibres are placed so that no overlap with aggregates occurs.
Consequently, the ITZs which are located at the interface between aggregates and matrix would not be expected to be strongly influenced by fibres.

\section{Conclusions}
Network meso-scale analyses of fracture processes of periodic cells subjected to direct tension were performed with the aim to investigate the link between material heterogeneity and width of the fracture process zone.
The meso-structures studied here consist of a quasi-brittle matrix with aggregates, fibres and combinations of aggregates and fibres.
For all material configurations, the width of the fracture process zone reduces abruptly after the peak load to the width of a rough crack.
This strong localisation happens very early in the post-peak regime at a stage at which very little energy has been dissipated during the fracture process.
For material configurations which include only matrix and aggregates, the width of the fracture process zone remains constant after the abrupt drop. 
For material configurations with fibres, the width of the fracture process zone increases after the drop since the slip between fibres and matrix contributes to the energy dissipation.

\section{Acknowledgements}
The numerical analyses were performed with the nonlinear analyses program OOFEM \cite{Pat12} extended by the present authors.

\bibliographystyle{kbib}
\bibliography{general}

\end{document}